\documentclass[preprint,review,12pt]{elsarticle}

\usepackage{amssymb}
\usepackage{amsthm}
\usepackage{amsmath}
\usepackage{multirow}
\usepackage{booktabs}
\usepackage{graphicx}
\usepackage{float}
\usepackage{url}
\usepackage{epstopdf}
\usepackage{pdflscape}
\usepackage{amsthm}
\usepackage{amsmath}
\usepackage[english]{babel}
\usepackage{subfigure}
\usepackage{threeparttable}

\usepackage{algorithm,algorithmicx,algpseudocode}
\usepackage{setspace}

\usepackage{xcolor}

\usepackage{tabularx}
\usepackage{graphicx}
\usepackage{caption}
\usepackage{threeparttable}
\usepackage[figuresright]{rotating}
\usepackage[toc,page,title,titletoc,header]{appendix} 

\makeatletter

\newcommand{\Rmnum}[1]{\expandafter\@slowromancap\romannumeral #1@}
\makeatother

\journal{XXX}

\begin{document}
\begin{frontmatter}
\title{A Feature-aware SPH for Isotropic Unstructured Mesh Generation}

\author[1]{Zhe Ji}
\ead{zhe.ji@tum.de}

\author[2]{Lin Fu}
\ead{linfu@stanford.edu}

\author[1]{Xiangyu Hu}
\ead{xiangyu.hu@tum.de}

\author[1]{Nikolaus Adams}
\ead{nikolaus.adams@tum.de}

\address[1]{Chair of Aerodynamics and Fluid Mechanics, Department of Mechanical Engineering, Technical University of Munich, 85748 Garching, Germany}

\address[2]{Center for Turbulence Research, Stanford University, Stanford, CA 94305, USA}

\begin{abstract}
\label{S:Abstract}
	In this paper, we present a feature-aware SPH method for the concurrent and automated isotropic unstructured mesh generation. Two additional objectives are achieved with the proposed method compared to the original SPH-based mesh generator (Fu et al., 2019). First, a feature boundary correction term is introduced to address the issue of incomplete kernel support at the boundary vicinity. The mesh generation of feature curves, feature surfaces and volumes can be handled concurrently without explicitly following a dimensional sequence. Second, a two-phase model is proposed to characterize the mesh-generation procedure by a feature-size-adaptation phase and a mesh-quality-optimization phase. By proposing a new error measurement criterion and an adaptive control system with two sets of simulation parameters, the objectives of faster feature-size adaptation and local mesh-quality improvement are merged into a consistent framework. The proposed method is validated with a set of 2D and 3D numerical tests with different complexities and scales. The results demonstrate that high-quality meshes are generated with a significant speedup of convergence.

\end{abstract}


\end{frontmatter}

\section{Introduction}
\label{S:introduction}

	Automated mesh generation is a critical and challenging topic for a wide range of scientific problems. In recent years, tremendous advancements have been made especially in unstructured triangular and tetrahedral mesh generation methods \cite{du2006recent}. The most well-established unstructured mesh generation methods can be classified into five categories, i.e. advancing front methods (AFT) \cite{schoberl1997netgen}\cite{lohner2014recent}, Octree refinement-based methods \cite{shephard1991automatic}\cite{yerry1984automatic}, Delaunay refinement-based methods \cite{shewchuk2002delaunay}\cite{chew1997guaranteed}, Delaunay variational-based methods \cite{ni2017sliver}\cite{du1999centroidal} and Particle-based methods \cite{FU2019396}\cite{zhong2013particle}\cite{bronson2010particle}. In addition to the aforementioned methods, lots of hybrids exist too especially in the context of parallel mesh generation \cite{chrisochoides2006parallel}. For a complete review of mesh generation methods, we refer to \cite{chrisochoides2006parallel}\cite{owen1998survey}\cite{frey2007mesh}.

	There are generally two steps for most mesh generators to obtain the final mesh, i.e. the initial mesh generation step and the mesh-quality improving step, providing the input geometry and the target feature-size function. It is uncommon that any mesh generation package can achieve both the objectives of target feature-size distribution and desired mesh quality without some form of mesh-quality improvement procedure \cite{owen1998survey}. The mesh-improvement algorithms mostly are different from the algorithms used for obtaining the initial mesh. For example, the code CGALmesh \cite{jamin2015cgalmesh} employs the restricted Delaunay refinement approach for the initial mesh generating. Then different types of optimization methods, e.g. Optimal Delaunay Triangulation \cite{chen2004optimal} and vertex perturbation \cite{tournois2009perturbing}, can be applied or combined sequentially to improve the mesh quality. Moreover, the computational time consumptions of the two steps are different too. Due to the rapid development of parallel algorithms, the current state-of-the-art mesh generators \cite{soner2015generating}\cite{feng2016two} can generate billions of meshes within minutes. However, the mesh-improvement step can be several orders of magnitude more expensive than the sequential meshing time if a high threshold is set for mesh quality and if the geometry is complicated \cite{klingner2008aggressive}\cite{chen2017domain}. Therefore, the two-step procedure hinders the automation of high-quality mesh generation. Meanwhile, developing the parallel version of different mesh-improving algorithms increases the workload and difficulties for software maintenance.

	Moreover, in order to generate an initial mesh, a certain sequence is required for most mesh generators. For AFT, the initial front has to be meshed first before being advanced, e.g. in order to generate a volumetric mesh, the surfaces of input geometry or sub-domain boundaries have to be meshed first \cite{lohner2014recent}. For Delaunay refinement-based methods, a surface mesh is required too as the input for conforming boundary tetrahedralization. Subsequently, the volumetric mesh is generated with interior Delaunay refinement \cite{du2003tetrahedral}. The same process applies to generating a surface mesh, i.e. the boundaries of the surfaces have to be meshed first \cite{meyer2008particle}. For the particle-based method \cite{yamakawa2000high}\cite{bronson2010particle}\cite{FU2019396}, the sampling and optimization of particle positions are also proceeded in a hierarchical way. The quality of the mesh at feature boundaries has significant influence on the resulting interior mesh \cite{frey1996delaunay}. For complex 3D geometries, obtaining a high-quality surface mesh is non-trivial, which results in even more time for generating the volumetric mesh. In this sense, concurrent mesh generation independent of feature type and dimension for the above-mentioned methods is not well developed. The concurrency issue is mitigated in a parallel environment, where the geometry can be divided into sub-domains and meshed separately \cite{chrisochoides2006parallel}\cite{feng2018hybrid}. In \cite{feng2016two}, a hybrid two-level Locality-Aware Parallel Delaunay imaging-to-mesh conversion algorithm (LAPD) is developed to exploit the parallelism from both coarse- and fine-grained perspectives. Although partitioning the geometry into sub-domains allows for better concurrency, inside each local mesh generator the same steps have to be executed in sequential.

	To tackle the aforementioned issues, several critical properties, e.g. inherently suitable for parallel computing, independent of geometry features, consistent for initial mesh generation and mesh-quality optimization, guaranteed convergence and etc., are required for the mesh generator. In this paper, the particle-based mesh generation methods are focused to achieve the automated and concurrent unstructured mesh generation due to their unique characteristics.

	The particle-based mesh generation methods share high similarity with the Delaunay variational-based methods. Both approaches require a target density/energy function \cite{ni2017sliver}\cite{du2003tetrahedral}\cite{FU2019396}. The target mesh-vertex distribution is calculated directly from the target density/energy function. The mesh quality is iteratively improved by applying different numerical schemes and by minimizing the energy or interpolation error. The key difference between the two approaches is whether the connectivity information is required during the optimization procedure \cite{ji2019consistent}. For particle-based methods \cite{FU2019396}\cite{zhong2013particle}\cite{meyer2005robust}, the triangulation/tetrahedralization of the mesh vertices is only applied once the mesh is generated, and the mesh quality is improved implicitly by pair-wise particle forces. While for Delaunay variational-based methods, re-triangulation/re-tetrahedralization is required every iteration to determine a new position for each vertex \cite{ni2017sliver}. Another appealing characteristic of the particle-based methods is that owing to the Lagrangian nature, the method is inherently suitable for large-scale parallel computing and can achieve scalable performance \cite{ji2018new}. Moreover, due to the simplicity of the method, it is easy to program and maintain with various parallel techniques, e.g. Message Passing Interface (MPI) \cite{MPI}, OpenMP \cite{openmp08} and CUDA \cite{Nickolls:2008:SPP:1365490.1365500}. Several well-established particle-based codes are already available for various architectures \cite{crespo2015dualsphysics}\cite{plimpton1995fast}\cite{incardona2019openfpm}.

	Previously, we have developed an unstructured mesh generator based on Lagrangian-particle fluid relaxation strategy \cite{FU2019396}, and further extended to adaptive anisotropic mesh generation \cite{FU2019396AN}. The target-density function is defined on a multi-resolution Cartesian background mesh considering the effects of distance to the geometry surface, curvature and singularity points. A set of physics-motivated governing equations has been then proposed and solved by adaptive-smoothing-length Smoothed Particle Hydrodynamics (SPH) \cite{monaghan1992smoothed}. By introducing a tailored equation-of-state (EOS), the relative discrepancy of particle density and target density is characterized as pseudo pressure. The pressure gradient results in pair-wise particle interaction force and drives particles towards target density distribution while maintaining a regularized and isotropic distribution \cite{ji2019consistent}. Later, the method is extended to parallel environment utilizing both MPI and Thread Building Blocks (TBB) \cite{contreras2008characterizing}\cite{fu2017novel} techniques. By introducing a repulsive surface tension model between distinct partitioning sub-domains, the targets of domain decomposition, communication volume optimization and high-quality unstructured mesh generation are achieved simultaneously within the same framework.

	However, the method still suffers from several problems. First, due to the friction model, particle velocities are nullified every timestep to maintain numerical stability. Therefore, a large number of iterations is needed to achieve a convergence \cite{FU2019396}. Although this issue can be mitigated by starting from an initial particle configuration generated from another mesh generator, the method then functions only as a mesh-improving algorithm. Second, since particles at geometry boundaries are treated as boundary conditions of inner particles and no special treatments are applied to remedy the lack of kernel support at the boundary vicinity, a well distributed particle configuration is required in advance to provide sufficient support. Therefore, a serial sequence is required to generate the final mesh \cite{FU2019396}.

	In this paper, we propose an improved particle-based mesh generator by developing a feature-aware SPH method to achieve two additional objectives in the same framework. First, by introducing a feature boundary correction term to address the lack of kernel support, particle interactions at the boundary vicinity are more consistent and accurate. If the target density function is smooth, particle evolution on feature curves, feature surfaces and volumes can be proceeded simultaneously without explicitly following a dimensional sequence. Second, a two-phase model is proposed to accelerate convergence of the particle evolution. Due to the feature boundary correction term, the constraint of zero velocities at each timestep can be significantly relaxed. Therefore the target of feature-size-adaptation can be achieved with much less iterations compared to the original algorithm. However, the mesh quality can be affected by high-frequency acoustic waves if momentum accumulation is allowed. The original simulation setup is preferred for final global mesh-quality improving. In this scenario, we propose to characterize our mesh-generation procedure by a feature-size-adaptation phase and a mesh-quality-optimization phase. Different simulation parameters are utilized for each phase. Consequently, the initial mesh generation and mesh-improvement steps are merged into a monolithic formulation.

	The paper is arranged as follows: In Section \ref{S:an_over_view}, the SPH-based isotropic mesh generator developed in \cite{FU2019396} is first briefly reviewed. Then the proposed feature-aware SPH method is introduced in detail in Section \ref{S:numerical_methods}. The feature definition, the correction term for feature boundaries and the two-phase mesh generation model are elaborated respectively. Initial particle sampling, particle stabilization strategy, flowchart and etc. are presented in this section too. In Section \ref{S:validation}, a set of validation tests are carried out considering both triangular surface-mesh and tetrahedral volumetric-mesh generation with the presence of various sharp features (creases, sharp edges and singularity points). Conclusion remarks in terms of the performance of the proposed method are given in the last section.

\section{The SPH-based isotropic unstructured mesh generator}
\label{S:an_over_view}

	In this section, the SPH-based isotropic mesh generator developed in \cite{FU2019396} is briefly reviewed.

	\subsection{Geometry definition}
	\label{S:geometry_definition}

		The surface of the underlying geometry is represented by a zero level-set function \cite{osher1988fronts}
		\begin{equation}
		\label{eq:level_set_function}
			\Gamma=\{(x,y)|\phi(x,y,t)=0\}.
		\end{equation}
		The positive phase, i.e. $\Gamma_{+}=\{(x,y)|\phi(x,y,t)>0\}$, is defined as the volumetric mesh-generation region. The zero level-set, i.e. $\Gamma_{0}=\{(x,y)|\phi(x,y,t)=0\}$, is defined as the surface mesh-generation region. A multi-resolution block-structured Cartesian background mesh is employed to discretize the level-set function.

	\subsection{Target feature-size definition}
	\label{S:target_feature_size_definition}

		Since the underlying mesh generation method falls into the category of variation-based approaches, a target feature-size function is required for the optimization of both mesh-vertices distribution and mesh quality. In general, the target feature-size function $h_t$ can be defined considering arbitrary characteristic fields. In \cite{FU2019396}, the effects of distance to the geometry surface $\phi$, the minimum distance from the geometry singularities $\psi$ and the diffused curvature field in the domain $\kappa$ are considered,
		\begin{equation}
		\label{eq:target_mesh_gen}
			\left\{
				\begin{array}{cr}
					h_t   =f(\phi, \psi, \kappa), & \\
					\rho_t=g(\phi, \psi, \kappa), & 
				\end{array}
			\right.
		\end{equation}
		where $\rho_t$ is the target density function defined for the relaxation procedure. $\rho_t$ can be obtained from $h_t$ following $\rho_t=\frac{1}{h^d}$, where $d$ is the spatial dimension. The three characteristic fields contributing to the feature-size function are calculated by solving three modeling equations, utilizing the same Cartesian background mesh. 

		Based on $\rho_t$, the total mass for generating a volume mesh defined in domain $\Omega$ and the total mass for generating a surface mesh can be calculated by integrating $\rho_t$ over the positive-phase of the geometry and the geometry surface respectively. For details, we refer to \cite{FU2019396}.


	\subsection{Modeling equations and numerical discretization}
	\label{S:modeling_equations_and_numerical_discretization}

		The evolution of mesh vertices is computed based on a fluid relaxation process and is modeled as an isothermal compressible flow. The Lagrangian formation of the governing equations is
		\begin{equation}
		\label{eq:continuity_equation}
			\dfrac{d\rho}{dt}=-\rho\bigtriangledown\cdot\textbf{v},
		\end{equation}
		\begin{equation}
		\label{eq:momentum_equation}
			\dfrac{d\textbf{v}}{dt}=-\textbf{F}_p+\textbf{F}_v,
		\end{equation}
		\begin{equation}
		\label{eq:moving_equation}
			\dfrac{d\textbf{x}}{dt}=\textbf{v},
		\end{equation}
		where $\rho$ denotes the density, $\textbf{v}$ the velocity vector, $\textbf{x}$ the position. $\textbf{F}_p$ and $\textbf{F}_v$ denote the pressure force and the viscous force respectively.

		To close the system, an EOS is required
		\begin{equation}
		\label{eq:EOS}
			p=P_0(\dfrac{\rho}{\rho_t})^{\gamma},
		\end{equation}
		where $P_0$ is a constant pressure, and $\gamma > 0$ is a user-defined parameter. This EOS drives particles to relax to the target distribution. Once an equilibrium state has been reached pressure becomes constant across the computational domain, and consequently the objectives of achieving target mesh-vertices distribution and optimizing mesh quality are achieved simultaneously.

		The model equations can be discretized and solved by the Smoothed Particle Hydrodynamics method. By assuming $\gamma=2$ in the EOS, the momentum equation can be discretized as 
		\begin{equation}
		\label{eq:momentum_equation_discretized}
			\dfrac{d\textbf{v}}{dt}=-\sum_jm_j\Big(\dfrac{p_0}{\rho^2_{t,i}}+\dfrac{p_0}{\rho^2_{t,j}}\Big)\dfrac{\partial W(r_{ij},h_{ij})}{\partial r_{ij}}\textbf{e}_{ij}+\sum_jm_j\dfrac{2\eta_i\eta_j}{\eta_i+\eta_j}\Big(\dfrac{1}{\rho^2_{t,i}}+\dfrac{1}{\rho^2_{t,j}}\Big)\dfrac{\partial W(r_{ij},h_{ij})}{\partial r_{ij}}\dfrac{\textbf{v}_{ij}}{r_{ij}},
		\end{equation}
		where $h$ is the smoothing length and characterizes the interaction range of the kernel function, $W(r_{ij},h_i)$ is the kernel function, $\frac{\partial W(r_{ij},h_{ij})}{\partial r_{ij}}$ the derivative of kernel, $\textbf{r}_{ij}=\textbf{r}_i-\textbf{r}_j$ the connecting vector between particle $i$ and $j$, $\textbf{e}_{ij}=\frac{\textbf{r}_{ij}}{r_{ij}}$ the unit vector of $\textbf{r}_{ij}$, $\textbf{v}_{ij}=\textbf{v}_{i}-\textbf{v}_{j}$, $h_{ij}=\frac{{h_i+h_j}}{2}$ the averaged smoothing length of particle $i$ and $j$, and $\eta=\rho\nu$ the dynamic viscosity. By setting
		\begin{equation}
		\label{eq:viscosity_coefficient}
			\nu\sim0.1r_c|\textbf{v}|,
		\end{equation}
		where $r_c$ is the cut-off radius of particle interaction range, the local Reynolds number of the simulation is always on the order of $O(10)$. Meanwhile, in order to maintain the stability of the simulation, the velocities of particles are nullified to damp the kinetic energy after every timestep.


		The system is advanced in time using a simplified Velocity-Verlet scheme. The timestep size is calculated considering the CFL criterion, the viscous criterion, and the body force criterion,
		\begin{equation}
		\label{eq:time_step_size}
			\Delta t = \min\Big(0.25\sqrt{\frac{r_c}{|\textbf{a}|}},\frac{1}{40}\frac{r_c}{|\textbf{v}|},0.125\frac{r_c^2}{\nu}\Big),
		\end{equation}
		where the artificial speed of sound is assumed as $c_s\sim40|\textbf{v}|_{max}$.

\section{The feature-aware SPH method}
\label{S:numerical_methods}

	\subsection{Geometry definition}
	\label{S:fa_geometry_definition}

		Following \cite{FU2019396}, the zero level-set function is utilized for geometry definition due to the flexibility of defining complex geometries. A Cartesian background mesh is utilized to discretize the level-set function. The size of the background mesh is defined according to the minimum and maximum target-feature-size. In the proposed method, we assume that the minimum and maximum mesh size, i.e. target-feature-size, is known before generating a mesh. In order to maintain accuracy, we set the minimum grid size slightly smaller than the minimum target-mesh size. In this paper, we generally ensure that the minimum mesh size is 1.5 to 2 times of the background grid size. 

		For complex geometries, multiple sharp edges and singularity points may exist, which cannot be resolved by a single level-set function. Additional inputs are employed in this paper to define these features to facilitate the geometry recovery. Following \cite{FU2019396}\cite{ji2019consistent}, the positions of singularity points are directly imported and singularity particles are generated accordingly. Moreover, sharp edges are defined with piecewise-linear B-splines, and each sharp edge is assigned with a unique index. To cooperate with the feature-definition system, each segment of the curves is mapped onto the background grid and all the cells containing feature-curves are characterized as feature-cells.

	\subsection{Feature definition}
	\label{S:feature_definition}

\begin{figure}[H]
		  \centering
		    \includegraphics[width=0.8\textwidth]{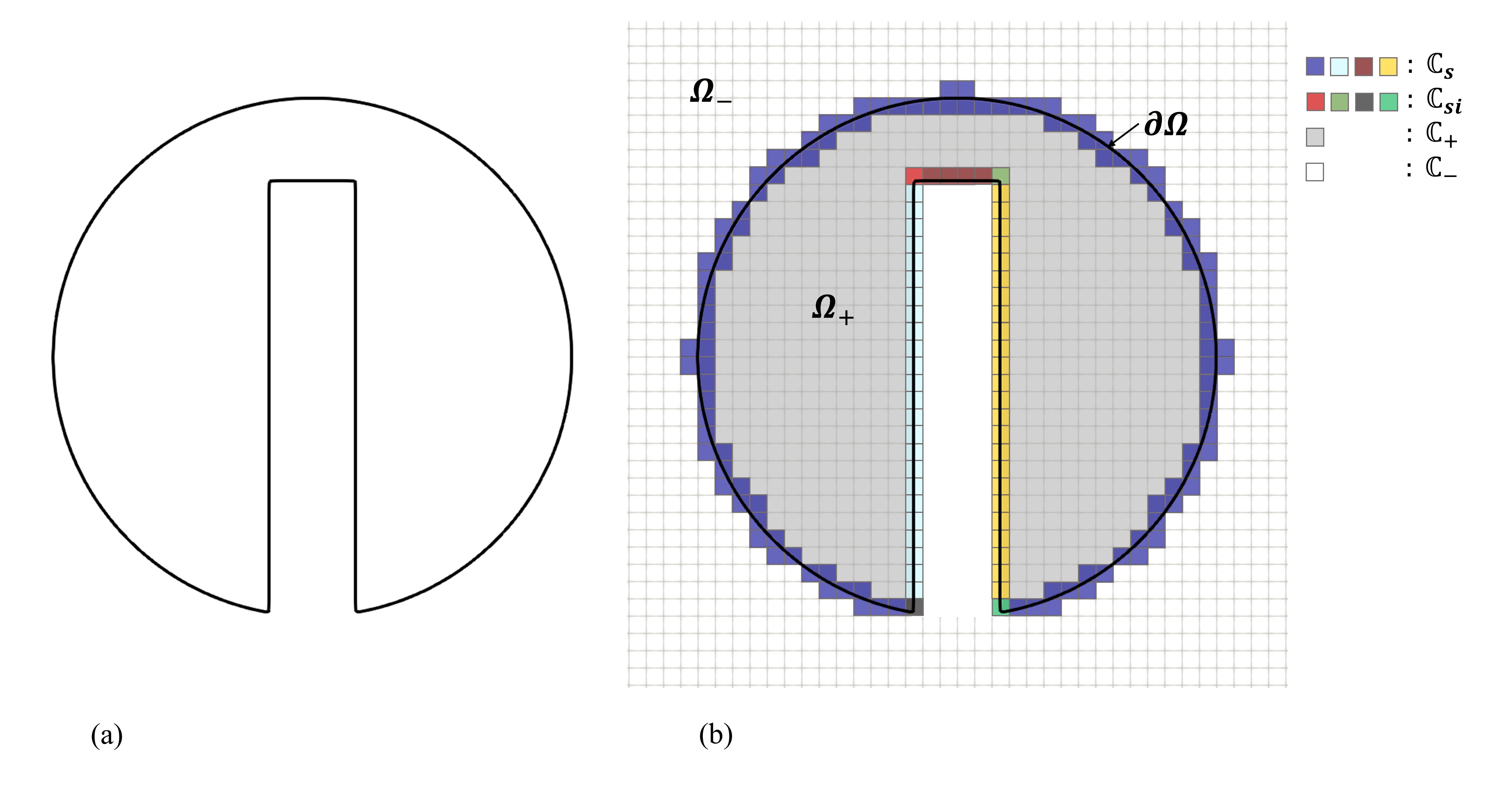}
		  \caption{(a) Contour of $\phi=0$ for the Zalesak's disk. (b) Tag system built for characterizing the geometry features.}
		\label{fig:tag_systems}
		\end{figure}

		The same Cartesian background mesh used for the definition of the level-set function introduced in Section \ref{S:geometry_definition} is utilized. To characterize the features, each cell $\mathbb{C}_{i}$ is assigned with a unique type and five categories of feature cells are defined accordingly, i.e. positive cell ($\mathbb{C}_{+}$), negative cell ($\mathbb{C}_{-}$), feature-surface cell ($\mathbb{C}_{s}$), feature-curve cell ($\mathbb{C}_{c}$) and singularity cell ($\mathbb{C}_{si}$). Figure \ref{fig:tag_systems} shows a simple example of the tag system with a 2D Zalesak's disk. In this case, four different types of cells are defined. For 3D geometries, the feature-curve cell can be defined if the geometry also contains sharp edges.

		For complex input geometries consisting of multiple sharp edges and singularity points, the geometry surfaces/volumes may be split into several regions. Therefore, these features are further characterized by a unique index. For the 2D Zalesak's disk (Figure \ref{fig:tag_systems}), a total number of 9 features are defined, i.e. 4 feature-surfaces, 4 singularities and 1 positive volume. All indices are represented with a unique color. In the following, each type of feature cell is distinguished by a subscript $k$, i.e. $\mathbb{C}_{*,k}$. A lookup table is constructed to associate the feature index with the feature type.

		Based on the above defined tag system, we can associate SPH particles with feature types. Four types of particles, i.e. singularity particles ($\mathbb{P}_{si}$), feature-curve particles ($\mathbb{P}_{c}$), feature-surface particles ($\mathbb{P}_{s}$) and positive particles ($\mathbb{P}_{+}$), are defined. In addition, based on the feature cell SPH particles lie in, the same feature index can be assigned too to each particle according to the cell feature index, and the same subscript $k$ is used.

		To evaluate the total number of mesh vertices, the target density is integrated first in each feature region to obtain the total mass of each feature. For volume mesh, the total mass $M_{v,k}$ of feature volume $V_k$ is calculated by
		\begin{equation}
		\label{eq:total_mass_2}
			M_{v,k}=\int_{V_k}\rho_t dv,
		\end{equation}
		The total mass for generating a surface mesh ($M_{s,k}$) or line mesh ($M_{c,k}$) can be calculated similarly with
		\begin{equation}
		\label{eq:total_mass_3}
			\left\{
				\begin{array}{c}
					M_{s,k}=\int_{S_k}\rho_tds,\\
					M_{c,k}=\int_{C_k}\rho_tds.
				\end{array}
			\right.
		\end{equation}
		where $S_k$ and $C_k$ denote the feature surface and feature curve with index $k$ respectively. The calculation procedure is completely parallelized since we can evaluate the mass in each cell independently. Only a reduction operation is required at the end for cells of the same feature index. The total number of mesh vertices, i.e. SPH particles, can then be calculated assuming each particle possesses unit mass following \cite{FU2019396}\cite{ji2019consistent}.

	\subsection{Correction term for feature boundaries}
	\label{S:Weighted_SPH_for_feature_boundaries}

		During the mesh-generation process, particles belonging to different features are not mutually independent. In the current paper, the interaction of particles between different features can be specified once the above-mentioned tag system is constructed. Similarly to \cite{FU2019396}\cite{ji2019consistent}, the interaction relationship is characterized in a hierarchical way following the dimensional sequence, e.g. $\mathbb{P}_{c}$ provide repulsive force for $\mathbb{P}_{s}$ and $\mathbb{P}_{+}$ to prevent penetration and $\mathbb{P}_{si}$ are used as boundary conditions for all the other types of particles. Moreover, for particles with the same feature type but assigned with different feature indices, no interactions are applied and they are evolved independently.

\begin{figure}[H]
		  \centering
		    \includegraphics[width=0.6\textwidth]{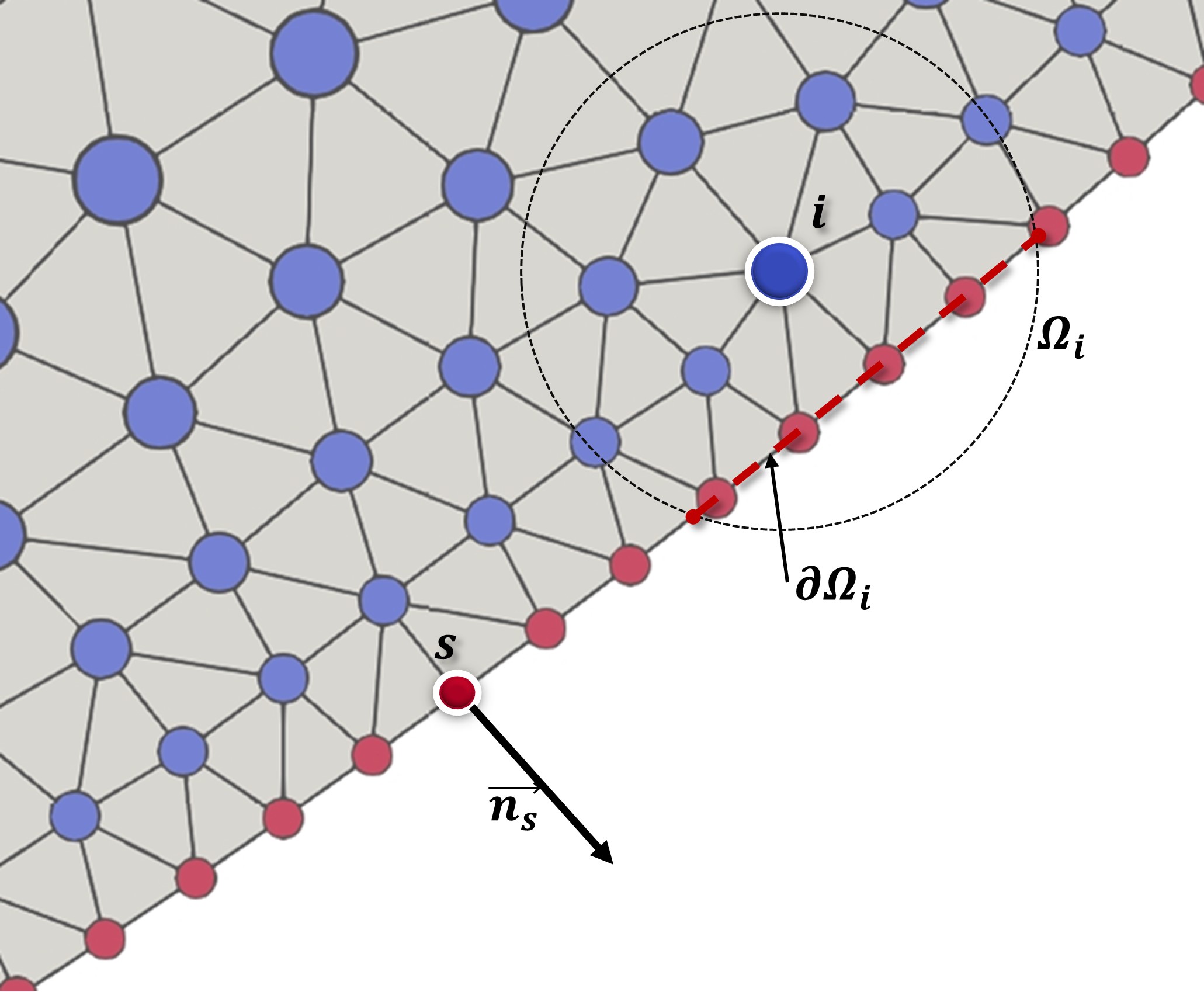}
		  \caption{Interaction with boundary particles and illustration of inconsistency in kernel support region.}
		\label{fig:BC_term}
		\end{figure}

		Fig. \ref{fig:BC_term} shows the interaction relationship between $\mathbb{P}_{s}$ (red circles) and $\mathbb{P}_{+}$ (blue circles) in a 2D scenario. In this case, $\mathbb{P}_{s}$ is used to impose the boundary conditions for $\mathbb{P}_{+}$. For particle $i$ (the highlighted blue circle), all red particles within the cutoff range (the dotted circle) are treated as boundary particles to enforce the impermeability condition and to complete kernel support.

		There are different approaches to model the interactions between ``interior" particles and ``boundary" particles. Due to the meshless and Lagrangian nature of SPH, imposing of robust and accurate boundary conditions is difficult \cite{chiron2019fast}. In our previous work \cite{FU2019396}, no special treatment has been applied. In \cite{FU2019396AN}, additional ghost particles are generated and evolved dynamically at the boundaries to prevent particles from penetrating the domain boundary in order to generate adaptive anisotropic triangular meshes. Since more particles are needed to achieve the desired mesh quality, additional computational costs and memory are required, which is critical in 3D scenarios.

		In this paper, we propose to address this issue with a new approach, utilizing the geometry information from the level-set function. First, we recall that with the SPH method, the gradient of a field quantity $\psi$ as the function of the coordinate $\textbf{x}$ can be calculated through a convolution integral over the cutoff range $\Omega$ of a kernel function $W$ as
		\begin{equation}
		\label{eq:gradient_operator_traSPH}
			\bigtriangledown \psi(\textbf{x})= \int_{\Omega}\bigtriangledown\psi(\textbf{y}) W(\textbf{r},h)d\textbf{y}.
		\end{equation}
		A discrete approximation of Eq. \ref{eq:gradient_operator_traSPH} for a set of particles at positions $\textbf{r}_i$ is
		\begin{equation}
		\label{eq:gradient_operator_disc_traSPH}
			\bigtriangledown \psi_i = \sum_j (\psi_{i}+\psi_{j})\bigtriangledown_iW(\textbf{r}_{ij}, h_{ij})V_j,
		\end{equation}
		where the summation is performed over all the neighboring particle $j$ within the cutoff range of particle $i$ with weight $W(\textbf{r}_{ij}, h_{ij})$, and $V_j$ is the volume of particle $j$.

		Eq. \ref{eq:gradient_operator_traSPH} and Eq. \ref{eq:gradient_operator_disc_traSPH} are consistent only when the cutoff range of the kernel function is fully inside the domain \cite{amicarelli20133d}\cite{chiron2019fast}\cite{feldman2007dynamic}. For the situation illustrated in Fig. \ref{fig:BC_term}, this full kernel support condition is no longer valid for $\mathbb{P}_{+}$ near geometry boundaries. To reproduce the kernel approximation at the boundary vicinity, a renormalization term $\gamma(\textbf{x})$ can be introduced to Eq. \ref{eq:gradient_operator_traSPH} \cite{chiron2019fast} \cite{hermange20193d}
		\begin{equation}
		\label{eq:gradient_operator_renormSPH}
			\bigtriangledown \psi(\textbf{x})= \dfrac{1}{\gamma(\textbf{x})}\int_{\Omega}\bigtriangledown\psi(\textbf{y})W(\textbf{r},h)d\textbf{y},
		\end{equation}
		where
		\begin{equation}
		\label{eq:gamma_renormSPH}
			\gamma(\textbf{x})= \int_{\Omega}W(\textbf{r},h)d\textbf{y}.
		\end{equation}

		Upon integration by parts, one can obtain
		\begin{equation}
		\label{eq:gradient_operator_renormSPH_bypart}
			\bigtriangledown \psi(\textbf{x})= -\dfrac{1}{\gamma(\textbf{x})}\int_{\Omega}\psi(\textbf{y})\bigtriangledown W(\textbf{r},h)d\textbf{y} + \dfrac{1}{\gamma(\textbf{x})}\int_{\partial\Omega}\psi(\textbf{y})W(\textbf{r},h)\textbf{n}dy,
		\end{equation}
		where $\partial\Omega$ is the boundary of $\Omega$ and $\textbf{n}$ is the normal direction of $\partial\Omega$ \cite{chiron2019fast} (see Fig. \ref{fig:BC_term}). Several approaches can be applied to discretize the equation based on how the geometry boundary is characterized, i.e. by particles or discrete segments \cite{hermange20193d}. In this paper, since we already use feature-surface, feature-curve and singularity particles to discretize the geometry boundaries, the fully discretized form of the gradient operator is employed,
		\begin{equation}
		\label{eq:gradient_operator_disc_renormSPH}
			\bigtriangledown \psi_i = -\dfrac{1}{\gamma_{i}}\sum_{j} (\psi_{i}+\psi_{j})\bigtriangledown_iW(\textbf{r}_{ij}, h_{ij})V_j + \dfrac{1}{\gamma_{i}}\sum_{b} (\psi_{i}+\psi_{b})W(\textbf{r}_{ib}, h_{ib})\textbf{n}_{b}A_b,
		\end{equation}
		where the first summation is calculated for all particles $j$ with the same type and feature index, while the second summation is performed for all ``boundary" particles $b$ from the perspective of particle $i$. $\textbf{n}_{b}$ and $A_{b}$ denote the normal direction and ``area" of particle $b$, respectively.

		The renormalization term $\gamma(\textbf{x})$ of particle $i$ can be directly calculated similar to the discrete Shepard coefficient \cite{shepard1968two} with the target volume of particle $j$, i.e.
		\begin{equation}
		\label{eq:gamma_disc_renormSPH}
			\gamma_i= \sum_{j}W(\textbf{r}_{ij}, h_{i})V_{t,j},
		\end{equation}
		where the summation is carried out over all the neighboring particles of ``i", and $V_{t,j}=h_{t,j}^d$.

		Finally, we can rewrite the pressure force term in Eq. \ref{eq:momentum_equation_discretized} as
		\begin{equation}
		\label{eq:sph_pressure_force_disc_weighted}
			\textbf{F}_{p,i}=\dfrac{1}{\gamma_{i}}\sum_jm_j\Big(\dfrac{p_0}{\rho^2_{t,i}}+\dfrac{p_0}{\rho^2_{t,j}}\Big)\dfrac{\partial W(r_{ij},h_{ij})}{\partial r_{ij}}\textbf{e}_{ij}+\dfrac{1}{\gamma_{i}}\sum_{b} \Big(\dfrac{p_0}{\rho^2_{t,i}}+\dfrac{p_0}{\rho^2_{t,b}}\Big)W(\textbf{r}_{ib}, h_{ib})\textbf{n}_{b}A_{t,b},
		\end{equation}
		where the target information $A_{t,b}=h_{t,j}^{d-1}$ is used for simplicity.

		From a physical point of view, the extra terms introduced in the pressure force can be interpreted as a ``repulsive force" from the boundary particles, which can prevent positive-phase particles from penetrating to negative domain. Therefore, the system is more stable and robust. A regularized and converged particle distribution at the geometry boundary is not necessary before evolving particles inside the domain. The extra term in the momentum equation dynamically adjusts particle positions at the boundary vicinity and maintains the validity of impermeability boundary conditions. Thus the constraints of evolving the system following a hierarchical sequence can be eliminated, and less iterations are required for generating a mesh.

		Moreover, the implementation of the extra terms is straightforward. All the additional information can be directly calculated based on the level-set function and target feature-size function. The computational overhead of the proposed method is through the additional summation over all the neighboring particles before calculating the forces. A considerably small increase in computation-time is introduced while additional memory requirement is negligible. The performance will be assessed in Section \ref{S:validation}.


	\subsection{The two-phase mesh generation model}
	\label{S:The_two-phase_model}

		\subsubsection{Main idea}
		\label{S:Main_idea}

		Based on the proposed method in the previous subsection, the mesh-vertex placement procedure can be simplified so that particles of all feature types can be evolved concurrently. In this section, a two-phase mesh generation model is proposed to further improve convergence. 

		The key concept is to merge the initial mesh generation phase and mesh quality optimization phase into the same formulation with different set of simulation parameters. After initialization, particle velocities evolve in time to allow for accumulation of momentum. During this initial phase, particles are driven by relaxation forces and relax rapidly towards optimum positions (minimum residual). Owing to the proposed correction term in Section \ref{S:Weighted_SPH_for_feature_boundaries}, the system can maintain stability towards equilibrium state. Following this initial phase, particles are re-adjusted locally to improve final mesh quality. Re-adjustment is achieved by switching back to the original method \cite{FU2019396}\cite{ji2019consistent}, in order to suppress high-frequency acoustic noises and deterioration of resulting mesh quality.

		It is worth noting that, because an equilibrium state has already been achieved in the initial phase, particles are close to the target-feature-size distribution. Re-adjusting particle velocity at every iteration only locally affect the positions of mesh vertices. From the experiments in Section \ref{S:validation}, the target of mesh-quality improvement can be achieved within a relative small number of iterations.

		With the proposed model, the same governing equations are solved but with different set of parameters. No intermediate triangulation/tetrahedralization step is required, and the implementation is trivial. In the following subsections, technical details of the method are presented. 

		\subsubsection{The convergence error measurement}
		\label{S:The_adaptive_control_system}

		In order to find a suitable criterion to switch from the initial phase (Phase One) to the re-adjustment phase (Phase Two), a proper measurement of convergence to the targeted feature-size distribution is required. One possibility is to evaluate convergence by monitoring the total kinetic energy. However, in complex geometries with singularities, the gradient of the target function may not be smooth, which contributes to inter particle forces. Consequently, particle velocities never vanish in those cases and a proper criterion based on the kinetic energy is impractical.

		According to the definition of the EOS, upon reaching equilibrium, the smoothing length of each particle equals the target feature size, i.e. $h_i=h_{t,i}$, and remains constant. In \cite{hu2006multi}, the authors proposed to approximate the particle volume as
		\begin{equation}
		\label{eq:particle_volume}
			\widetilde{V}_i= \dfrac{1}{\sum_{j}W(\textbf{r}_{ij}, h_{i})},
		\end{equation}
		i.e. the inverse of the particle specific volume. Ideally, when the particle feature-size approaches the target, we can obtain
		\begin{equation}
		\label{eq:volume_to_target}
			\widetilde{V}_i \rightarrow \widetilde{V}_{t,i}=\dfrac{1}{\sum_{j}W(\textbf{r}_{ij}, h_{t,j})}.
		\end{equation}
		The calculation of $\widetilde{V}_i$ is less sensitive to the gradient of the target function and thus is more suitable as the measurement of convergence. The normalized form of the total ``volume" in each feature is
		\begin{equation}
		\label{eq:volume_integral}
			\{\overline{V}_{+,k}, \overline{V}_{s,k}, \overline{V}_{c,k}\} = \frac{1}{V_{ref,k}}\{\sum\limits_{i}^{i\in\mathbb{P}_{+,k}}\widetilde{V}_i, \sum\limits_{i}^{i\in\mathbb{P}_{s,k}}\widetilde{V}_i, \sum\limits_{i}^{i\in\mathbb{P}_{c,k}}\widetilde{V}_i\},
		\end{equation}
		where the notation ``$+$", ``$s$", ``$c$" denotes the feature type, $k$ the feature index, $V_{ref, k}$ the reference volume of feature $k$, witch is an approximated value of the total volume/area/length of the feature. We employ 1D/2D/3D SPH kernels for $\mathbb{P}_{c}$/$\mathbb{P}_{s}$/$\mathbb{P}_{+}$ to calculate $\widetilde{V}_i$ respectively.

		Based on Eq. \ref{eq:volume_integral}, we can obtain the time history of $\overline{V}_{*,k}$. If the relative error ($E_{t}$) between $\overline{V}_{*,k}$ and the last time-averaged (over a certain period) $\overline{V}_{*,k}$ is smaller then a threshold and the relative error ($E_{avg}$) between two consecutive time-averaged $\overline{V}_{*,k}$ also is smaller then a threshold, then we consider that the system has achieved the target of Phase One. The maximum value between $E_{t}$ and $E_{avg}$ is used as the measurements of the convergence error and is denoted as $\overline{E}_{sys}$. For all the cases in Section \ref{S:validation}, we use an interval of 200 timesteps for time averaging and $\overline{V}_{*,k}$ is measured every 20 timesteps, i.e. 10 sampling points each time-averaging period. The threshold is set to $5\times10^{-6}$.

		\subsubsection{Particle stabilization strategy}
		\label{S:Particle_stabilization_strategy}

			Although a correction term is introduced to address the issue of incomplete kernel support and to relax the need to setting particle velocity to zero each timestep, we set particle velocities to zero periodically to guarantee that particle distributions are regularized and that timestep sizes do not become very small. Extensive tests suggest that a period of 50-200 timesteps generally achieves a good balance between stability and convergence. In this paper, we use a period of 100 for all the cases below. In addition, we monitor timestep-size variations, when we observe a decrease by an order of magnitude compared to the previous step, particle velocities are set to zero. In some extreme situations, e.g. ill-posed initial particle sampling, multiple singularities, particle clustering during simulation, an additional damping term is introduced to the momentum equation during Phase One
			\begin{equation}
			\label{eq:momentum_equation_new}
				\dfrac{d\textbf{v}}{dt}=-\textbf{F}_p+\textbf{F}_v-\varepsilon\textbf{v},
			\end{equation}
			where $\varepsilon$ is the damping factor. In this paper, a range of $\varepsilon\in[0,0.2]$ is suggested. The damping term is not needed in Phase Two.

		\subsubsection{Remarks}
		\label{S:Simulation_parameters}

			In order to achieve the targeted feature-size adaptation and mesh-quality optimization with the same framework, a two-phase mesh generation model is proposed in this section. First, the convergence of the feature-size adaptation phase, i.e. Phase One, is monitored by introducing the normalized convolutional particle volume estimation term, i.e. $\overline{V}_{*,k}$. Once the convergence error of the simulation drops below a certain threshold, Phase One is completed. In Phase One, accumulation of particle momentum is allowed, which facilitates faster convergence to optimized mesh-vertex positions. In Phase Two, the original mesh generation scheme is employed, which suppresses high-frequency acoustic waves and achieves improved mesh-quality.

			The key differences between Phase One and Phase Two are two parameters, i.e. the period of velocity nullification and the damping factor. The suggested values/value ranges can be found in previous sections. Moreover, between Phase One and Phase Two, we use an intermediate period and an attenuation function to ramp down the parameters to ensure a smooth transition of the computation.

	\subsection{Feature-aware time marching}
	\label{S:Regional_timestepping}

		Particles are characterized according to feature types and feature indices. Forces between particles of different feature types are only one-sided. For particles of the same type but with different feature index, interactions are mutually exclusive. Consequently, we can evolve particles in groups, and each group possesses particles of identical feature index. The timestep size of each group can be different. Based on this observation, we modify Eq. \ref{eq:time_step_size} as
		\begin{equation}
		\label{eq:fa_time_step_size}
			\Delta t_{*,k} = \min\limits_{i\in\mathbb{P}_{*,k}}\Big(0.25\sqrt{\frac{r_{c,i}}{|\textbf{a}_i|}},\frac{1}{40}\frac{r_{c,i}}{|\textbf{v}_i|},0.125\frac{r_{c,i}^2}{\nu}\Big),
		\end{equation}
		where the notation ``$*$" represents different feature types.

		Based on the smooth target feature-size function assumption, in simple geometries the timestep sizes among all features should be similar. However, for complex geometries and large resolution variations, timestep size distribution across all feature types may deteriorate the convergence of the full system. By using feature-aware timestep sizes, asynchronous convergence of each feature improves robustness. Moreover, evolving particles in groups also facilitates pipeline parallelization of the code.

	\subsection{Initial particle sampling}
	\label{S:Initial_particle_sampling}

		In the previous particle-based mesh generation methods \cite{FU2019396}\cite{ji2019consistent}, the initial mesh vertices, i.e. SPH particles, are generated randomly with constant probability inside the computational domain. This approach demonstrates that the proposed methods can generate a mesh from extreme poor initial conditions. However, the iteration number required for obtaining a high-quality meshes from this initial condition is very high, thus making it impractical in realistic applications and much more expensive than other mesh generators \cite{frey2007mesh}\cite{ji2019consistent}.

		Instead of using constant probability function, the initial particle distribution is generated randomly with probabilities proportional to the target density field. We use the Mersenne Twister algorithm \cite{matsumoto1998mersenne} to calculate random numbers, which is of $O(1)$ time complexity. Therefore, no significant extra computational overhead is introduced. The initial sampling is closer to the target feature-size distribution, and the number of iterations to achieve an optimized mesh quality can be reduced.

	\subsection{Review of the algorithm}
	\label{S:Review_of_the_algorithm}

		A detailed flowchart of the proposed method is given as Algorithm \ref{alg:FeatureAwareMeshGen}. The method is implemented based on the parallel framework for adaptive-resolution SPH developed in \cite{ji2018new}\cite{linfu_fast_neighbor}. Since the equations of the proposed algorithm exhibit high similarities with standard SPH for gas dynamics, it is straightforward to implement the mesh generator from an existing SPH code. For more technical details we refer to \cite{ji2018new}\cite{ji2019lagrangian}.

		\begin{algorithm}[H]
		\footnotesize
		\setstretch{1.35}
		\caption{Flowchart of the feature-aware mesh generation method}
		\label{alg:FeatureAwareMeshGen}
		\begin{algorithmic}[1]
		\State Initialize the background Cartesian mesh and the level-set function (Eq. \ref{eq:level_set_function});
		\State Load user-defined feature curves and singularity points;
		\State Calculate the target density function (Eq. \ref{eq:target_mesh_gen}) based on the background mesh;
		\State Define the feature-based tag system for each cell ($\mathbb{C}_{i}$) and calculate the target information for mesh generation (Eq. \ref{eq:total_mass_2} and Eq. \ref{eq:total_mass_3});
		\State Allocate particles based on the target information and initial particle sampling;
		\While{$t<t_{end}$}
		  \State Update the multi-level data structure;
		  \Comment {\textit{See \cite{ji2018new} for detailed description}}
		  \For{$k = 0 \to K-1$}
		  \Comment {\textit{K is the total number of features}}
		    \If{$\overline{V}_{*,k}$ is not converged}
			  \State Define $\rho_t$, $h_t$, $\textbf{n}$ and etc. for $\mathbb{P}_{*,k}$ from background mesh;
			  \State Construct Verlet neighbor list;
			  \Comment {\textit{See \cite{ji2018new} for detailed description}}
			  \State Reset particle forces;
			  \State Reset particle velocities;
			  \Comment {\textit{See Section \ref{S:Particle_stabilization_strategy} for detailed description on when to do the velocity reinitialization}}
			  \State Calculate the renormalization term $\gamma_i$ (Eq. \ref{eq:gamma_disc_renormSPH})
			  \State Calculate pressure force $\textbf{F}_p$ (Eq. \ref{eq:sph_pressure_force_disc_weighted});
			  \State Map $\textbf{F}_p$ to feature $k$ for $\mathbb{P}_{*,k}\in(\mathbb{P}_s\cup\mathbb{P}_c)$ (Eq. 30 and Eq. 31 in \cite{ji2019consistent});
			  \State Calculate $\Delta t_{*,k}$ (Eq. \ref{eq:fa_time_step_size});
			  \State Update the mid-point velocity $\widetilde{\textbf{v}}_{n+\frac{1}{2}}$ (Eq. 26 in \cite{ji2019consistent});
			  \State Accumulate viscous force  $\textbf{F}_v$ (Eq. \ref{eq:momentum_equation_discretized});
  			  \State Map $\textbf{F}_v$ to feature $k$ for $\mathbb{P}_{*,k}\in(\mathbb{P}_s\cup\mathbb{P}_c)$ (Eq. 30 and Eq. 31 in \cite{ji2019consistent});
			  \State Calculate $\Delta t_{*,k}$ (Eq. \ref{eq:fa_time_step_size});
			  \State Update predicted velocity $\textbf{v}_{n+\frac{1}{2}}$ (Eq. 27 in \cite{ji2019consistent});
			  \State Update particle position $\textbf{r}_{n+1}$ (Eq. 28 in \cite{ji2019consistent});
			  \State Map $\textbf{r}_{n+1}$ to feature $k$ for $\mathbb{P}_{*,k}\in(\mathbb{P}_s\cup\mathbb{P}_c)$;
			  \Comment {\textit{See Section 3.5 in \cite{ji2019consistent}}}
			  \State Calculate $\overline{V}_{*,k}$ and check convergence;
			\EndIf
		  \EndFor
		  \If{Do post-processing}
		    \State Generate the corresponding mesh and calculate mesh quality; 
		  \EndIf
		\EndWhile
		\end{algorithmic}
		\end{algorithm}

\section{Validation and demonstration}
\label{S:validation}

	In this section, a set of numerical experiments varying from generating 2D triangular to 3D surface/volume meshes is considered. The performance of our method is assessed, and the results are compared with previous work and other established methods. In all the experiments below, we refer to the results obtained by the proposed method as ``\textit{improved}". All cases in this section are computed on a desktop workstation with Intel\textsuperscript{\textregistered} Xeon\textsuperscript{\textregistered} CPU (E5-2630V2, 2.60GHz, 12 threads).

	For 2D triangulation, the mesh is constructed following \cite{FU2019396} utilizing a local Voronoi tessellation. For 3D tetrahedralization, the open-source code TetGen \cite{si2015tetgen} is used. Flip operations included in TetGen (2-3 flip, 3-2 flip and 4-4 flip) are invoked to improve connectivity. It is worth noting that the mesh quality is only measured for comparison purposes. In the proposed method, the mesh quality is optimized implicitly and the triangulation/tetrahedralization step is not invoked in the mesh-generation procedure.

	The quality of isotropic triangular meshes is quantified by $G=2\sqrt{3}\frac{S}{PH}$ and the angle $\theta$, where $S$ is the triangle area, $P$ the half-perimeter and $H$ the length of the longest edge. $\theta_{max}$, $G_{avg}$, $G_{min}$/$\theta_{min}$ and $\theta_{min}^{\#}$ are the maximum, average, minimum values and the averaged minimum angle in each triangle respectively. $\theta_{<30}$ is the number of triangle with angle smaller than $30^{\circ}$. The quality of isotropic tetrahedral meshes is quantified by the dihedral angle $\theta$ and radius ratio $\gamma=3\frac{r_{in}}{r_{circ}}$, where $r_{in}$ is the inradius and $r_{circ}$ the circumradius of a tetrahedron. $\theta_{max}$, $\gamma_{avg}$, $\gamma_{min}$/$\theta_{min}$ and $\theta_{min}^{\#}$ are the maximum, average, minimum values and the averaged minimum dihedral angle in each tetrahedron respectively. The number of slivers are measured by counting the number of tetrahedra with different smallest dihedral angles, i.e. $10^{\circ}$, $20^{\circ}$, $30^{\circ}$ and $40^{\circ}$.

	\subsection{Square}
	\label{S:validation_square}

\begin{figure}[H]
		  \centering
		    \includegraphics[width=0.8\textwidth]{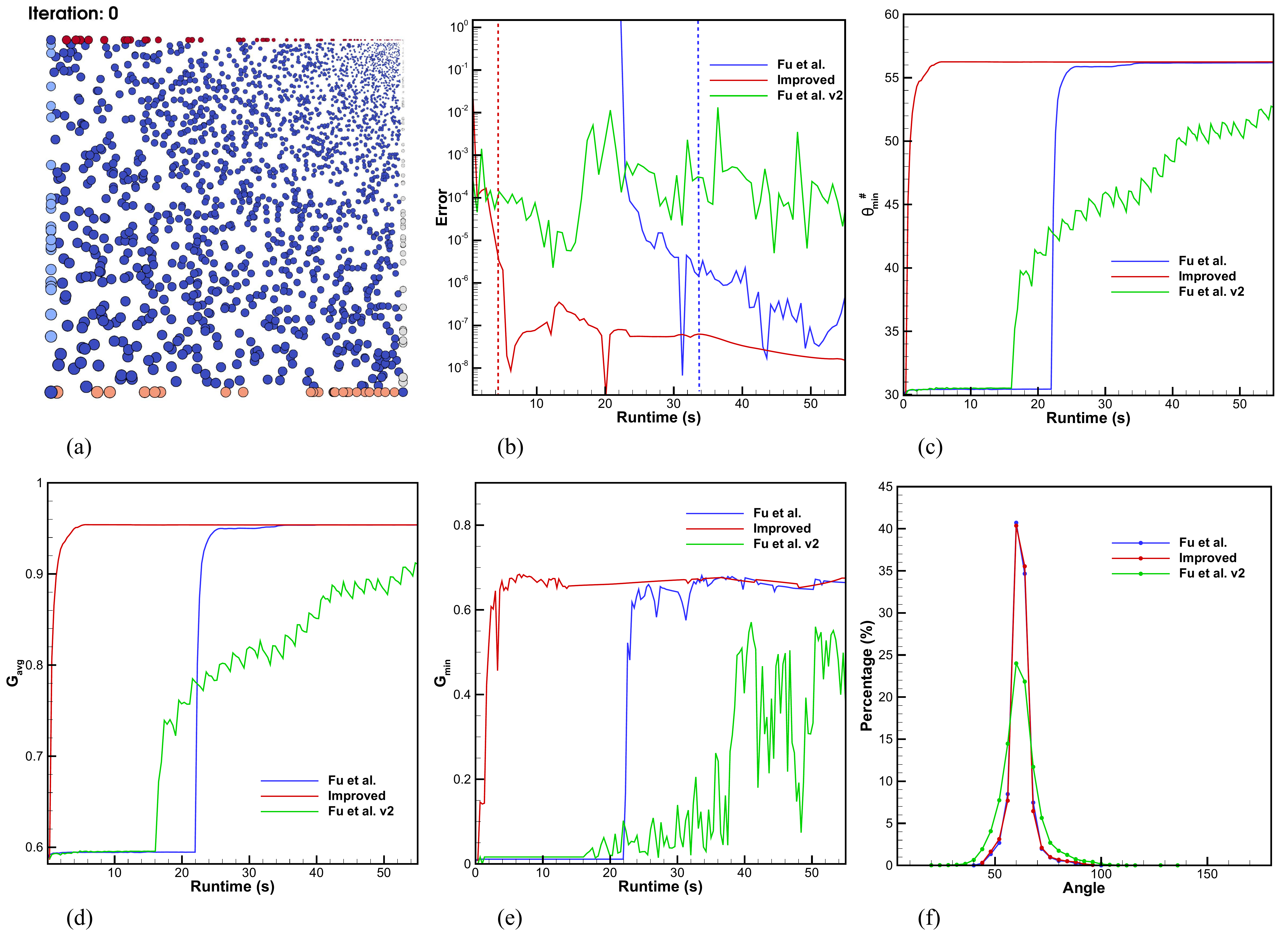}
		  \caption{Square: (a) Initial particle distribution. The convergence histories of (b) $\overline{E}_{sys}$, (c) $\theta_{min}^{\#}$, (d) $G_{avg}$ and (e) $G_{min}$. (f) Histogram of the angle distribution.}
		\label{fig:square_stats}
		\end{figure}

		First we consider a square with feature-size function 
		\begin{equation}
		\label{eq:h_t_square}
			h_t(x,y)=\frac{h_{max}-h_{min}}{100\sqrt{2}}\sqrt{(x-100)^2+(y-100)^2}+h_{min},
		\end{equation}
		where $(x,y)\in[0,100]$. The maximum and minimum feature-size ($h_{max}$ and $h_{min}$) are $4.88$ and $0.244$ respectively. The total number of particles is 2524. The results are compared with the previously proposed SPH-based mesh generator \cite{FU2019396}. A variant of the method \cite{FU2019396} is considered (referred to as [10.v2]), where particle velocities are set to zero with same period, i.e. every 100 timestep, as with the current method. All computations start with the same initial particle distribution (see Fig. \ref{fig:square_stats} (a)).

		The convergence history comparisons of $\overline{E}_{sys}$, $\theta_{min}^{\#}$, $G_{avg}$ and $G_{min}$ are shown vs. the total wall-clock runtime in Fig. \ref{fig:square_stats}. The angle distribution is measured and compared in Fig. \ref{fig:square_stats}. The dashed lines in Fig. \ref{fig:square_stats} (b) indicate that the system has achieved the prescribed error threshold of Phase One.

\begin{figure}[H]
		  \centering
		    \includegraphics[width=0.8\textwidth]{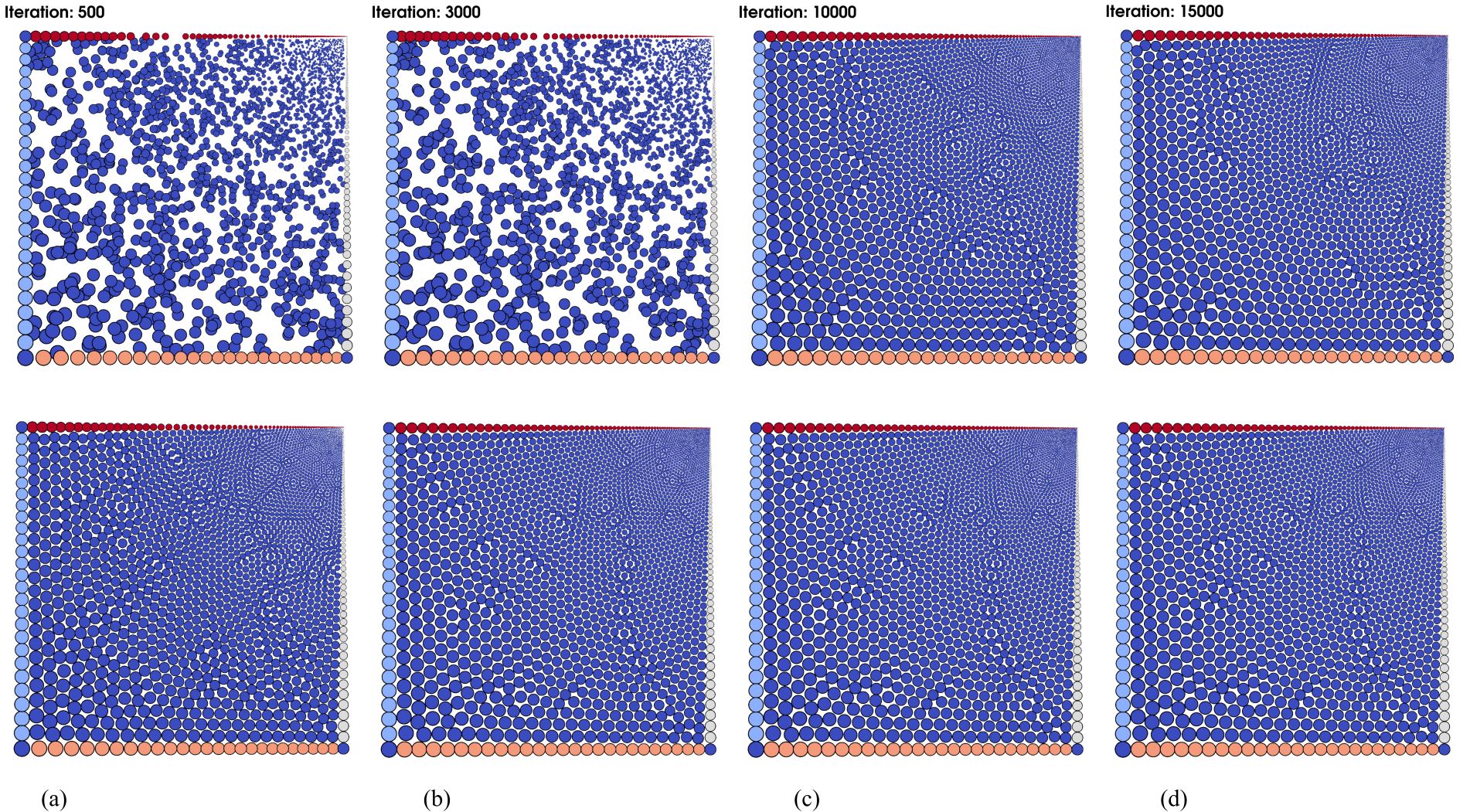}
		  \caption{Square: Particle distributions at (a) iteration 3000, (b) iteration 6000 and (c) iteration 12000. Upper row: results calculated by the algorithm developed by Fu et al. \cite{FU2019396}. Bottom row: results calculated by the proposed algorithm.}
		\label{fig:square_particle}
		\end{figure}

\begin{figure}[H]
		  \centering
		    \includegraphics[width=0.6\textwidth]{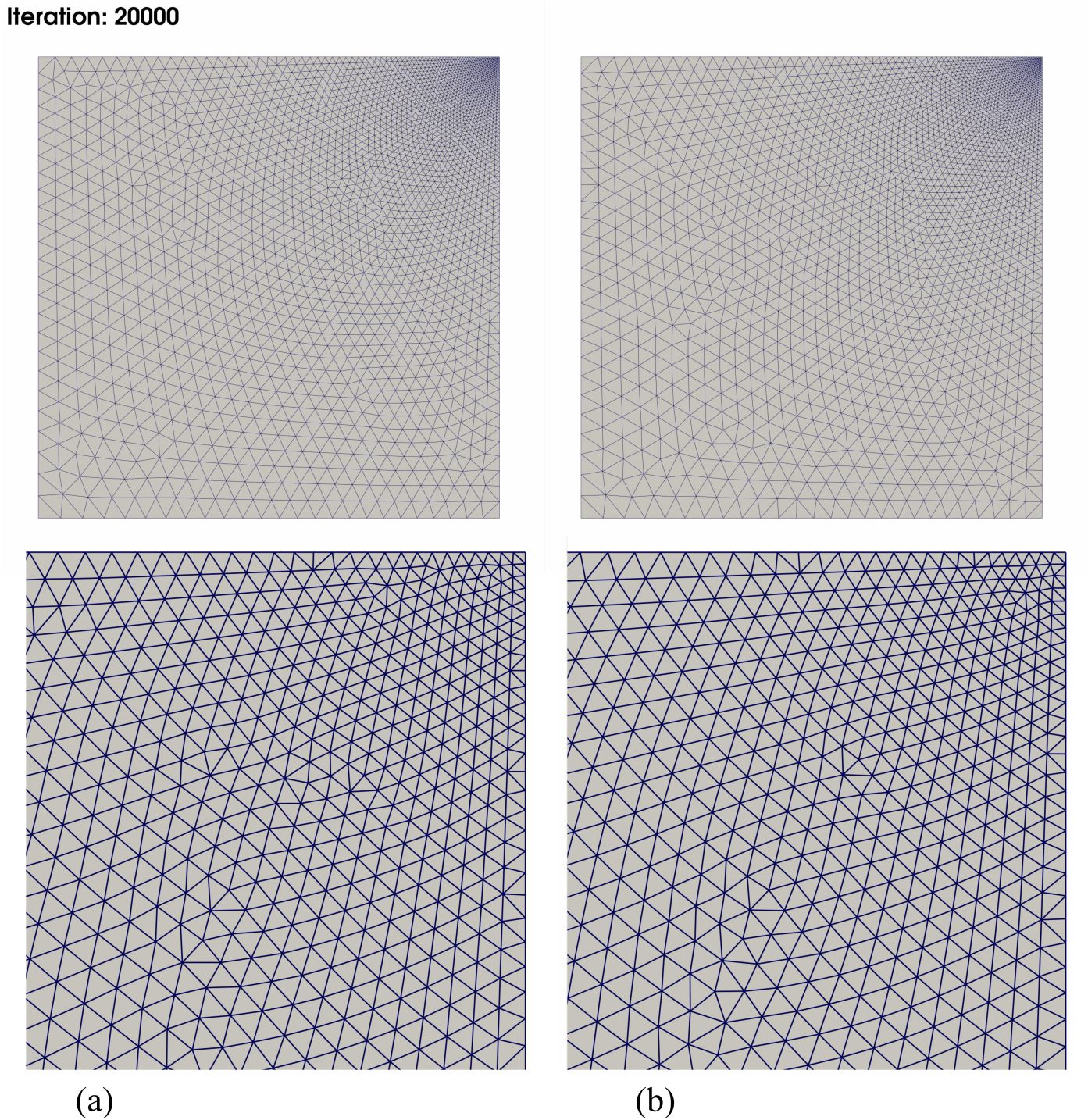}
		  \caption{Square: Delaunay triangulation of the final mesh and the zoomed-in view at iteration 20000. (a) Mesh generated by the algorithm developed by Fu et al. \cite{FU2019396}. (b) Mesh generated by the proposed algorithm.}
		\label{fig:square_mesh}
		\end{figure}

		First it is observed that [10.v2] fails to converge in the plotted time range, and the mesh-quality is the worst. Due to the absence of the feature-aware correction term proposed in this paper, the system is not able to maintain stability at the vicinity of the boundary when the accumulation of particle momentum is allowed. Successive particle losses at the domain boundary are observed. Remapping particle into the positive phase are required to handle the particle lost issue. Second, the ``\textit{improved}" method achieves the target-feature distribution significantly faster than the previous method \cite{FU2019396}. The runtime for both methods to achieve the target of Phase One are approximately $4.2s$ and $34s$, which implies a speedup factor of $8.1$. This can also be demonstrated by the particle distributions at different iterations (see Fig. \ref{fig:square_particle}). The quality of the generated meshes exhibits no obvious difference (see Fig. \ref{fig:square_stats} (c-f) and Table \ref{Tab:validation_square}), which can also be seen in the Delaunay triangulations of the final mesh in Fig. \ref{fig:square_mesh}.

		\begin{table}[h]
		\centering
		\caption{Mesh quality of the Square case}
		\scriptsize
		\label{Tab:validation_square}
		\newcommand{\tabincell}[2]{\begin{tabular}{@{}#1@{}}#2\end{tabular}}
		\begin{tabular}{>{\centering\arraybackslash}m{1.2cm}
		                >{\centering\arraybackslash}m{0.6cm}
		                >{\centering\arraybackslash}m{0.6cm}
		                >{\centering\arraybackslash}m{1cm}
		                >{\centering\arraybackslash}m{1cm}
		                >{\centering\arraybackslash}m{1cm}
		                >{\centering\arraybackslash}m{1cm}
		                >{\centering\arraybackslash}m{1cm}
		                >{\centering\arraybackslash}m{1cm}
		                >{\centering\arraybackslash}m{1cm}}
		\hline
		 & $G_{avg}$ & $G_{min}$ & $\theta_{max}$ & $\theta_{min}$ & $\theta_{min}^{\#}$ & $\theta_{<30}$ & $N_{tri}\footnotemark$ & $runtime$\footnotemark & $N_{iter}\footnotemark\ \footnotemark$\\ \hline
		 \textit{Fu et al.} & 0.95 &  0.66 &  96.20 & 38.38 & 56.13 & 0 & 4,831 & 34.82 & 12,200 \\
		 \textit{Improved}  & 0.95 &  0.67 &  94.85 & 40.11 & 56.23 & 0 & 4,828 & 5.07 & 1,600  \\
		\hline
		\end{tabular}
		\end{table}
		\footnotetext[1]{$N_{tri}$ the total number of triangles}
		\footnotetext[2]{$runtime$ denotes the total wall-clock time in seconds}
		\footnotetext[3]{$N_{iter}$ denotes the number of iterations for the measured runtime}
		\footnotetext[4]{All the mesh quality in this table are measured at $N_{iter}$, and remains consistent hereafter.}

	\subsection{Zalesak's disk (inside)}
	\label{S:validation_disk}

\begin{figure}[H]
		  \centering
		    \includegraphics[width=0.8\textwidth]{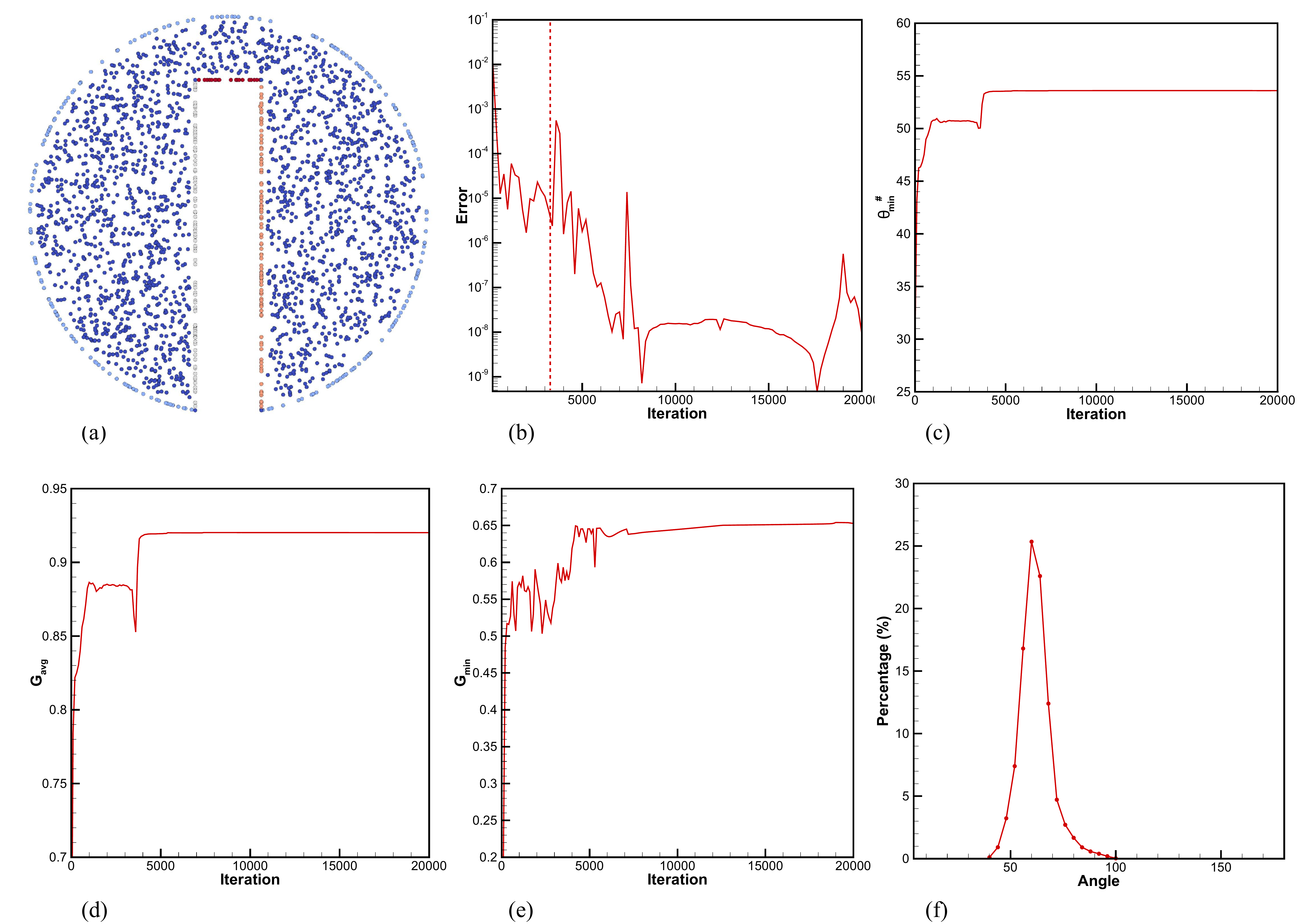}
		  \caption{Zalesak's disk:(a) Initial particle distribution. The convergence histories of (b) $\overline{E}_{sys}$, (c) $\theta_{min}^{\#}$, (d) $G_{avg}$, (e) $G_{min}$. (f) Histogram of the angle distribution.}
		\label{fig:disk_stats}
		\end{figure}

		Next, we consider the 2D Zalesak's disk following \cite{FU2019396}. The geometry consists of a slotted circle of radius 15. The width and length of the slot are 5 and 25 respectively. The mesh generation region is inside the slotted circle. The definition of feature-size function and $h_{min}$/$h_{max}$ are identical to \cite{FU2019396}. Instead of generating the initial sampling of particles following the target density function $\rho_t(\textbf{x})$, we use a uniform probability function to sample the particles (see Fig. \ref{fig:disk_stats} (a)), which is the same with \cite{FU2019396}.

\begin{figure}[H]
		  \centering
		    \includegraphics[width=0.8\textwidth]{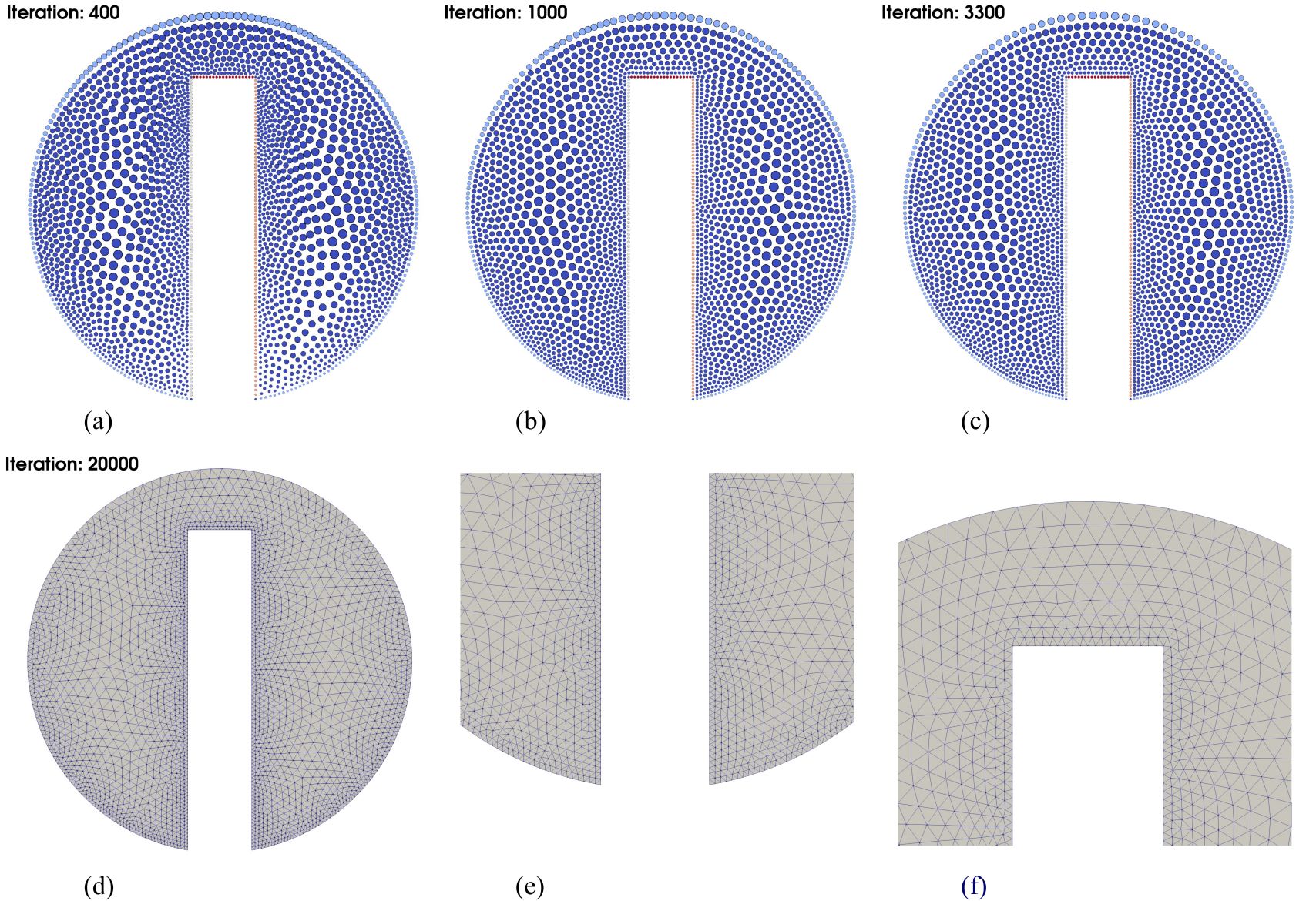}
		  \caption{Zalesak's disk: Particle distributions at (a) iteration 3000, (b) iteration 6000 and (c) iteration 12000. (d)(e)(f) Delaunay triangulation of the final mesh and the zoomed-in views at iteration 20000.}
		\label{fig:disk_particle}
		\end{figure}

		Based on the convergence histories of $\overline{E}_{sys}$, $\theta_{min}^{\#}$, $G_{avg}$ and $G_{min}$ in Fig \ref{fig:disk_stats} (b-c), a high-quality mesh is obtained and both the targets of Phase One and Phase Two are achieved. The system remains stable during the mesh-generation procedure, which is demonstrated by the particle distributions at different iterations (see Fig. \ref{fig:disk_particle}). The mesh quality calculated by the proposed method is close to the result from \cite{FU2019396}, and is compared in Table \ref{Tab:validation_disk}. The main differences of the two results are the iterations required to achieve the convergence. Since in \cite{FU2019396}, there is no explicit measurement of convergence, we consider the instant as converged when all the mesh-quality curves reach the plateau. The method \cite{FU2019396} achieves convergence after ca. 45,000 iterations, see on Fig. 9 of \cite{FU2019396}. For the proposed method, we obtain the same mesh-quality with only 4,200 iterations, i.e. 91\% reduction.

		\begin{table}[h]
		\centering
		\caption{Mesh quality of the Zalesak's disk case}
		\scriptsize
		\label{Tab:validation_disk}
		\newcommand{\tabincell}[2]{\begin{tabular}{@{}#1@{}}#2\end{tabular}}
		\begin{tabular}{>{\centering\arraybackslash}m{1.2cm}
		                >{\centering\arraybackslash}m{0.6cm}
		                >{\centering\arraybackslash}m{0.6cm}
		                >{\centering\arraybackslash}m{1cm}
		                >{\centering\arraybackslash}m{1cm}
		                >{\centering\arraybackslash}m{1cm}
		                >{\centering\arraybackslash}m{1cm}
		                >{\centering\arraybackslash}m{1cm}
		                >{\centering\arraybackslash}m{1cm}
		                >{\centering\arraybackslash}m{1cm}}
		\hline
		 & $G_{avg}$ & $G_{min}$ & $\theta_{max}$ & $\theta_{min}$ & $\theta_{min}^{\#}$ & $\theta_{<30}$ & $N_{tri}$ & $runtime$ & $N_{iter}$\\ \hline
		 \textit{Improved}  & 0.92 &  0.65 &  97.73 & 37.23 & 53.50 & 0 & 4,551 & 15.00 & 4200  \\
		 \textit{Fu et al.} & 0.92 &  0.66 &  ---   & 28.81 & 53.89 & 0 & ---   & $\sim$400 & $\sim$45K  \\
		\hline
		\end{tabular}
		\end{table}

	\subsection{Tyrannosaurus rex}
	\label{S:validation_disk}

\begin{figure}[H]
		  \centering
		    \includegraphics[width=0.8\textwidth]{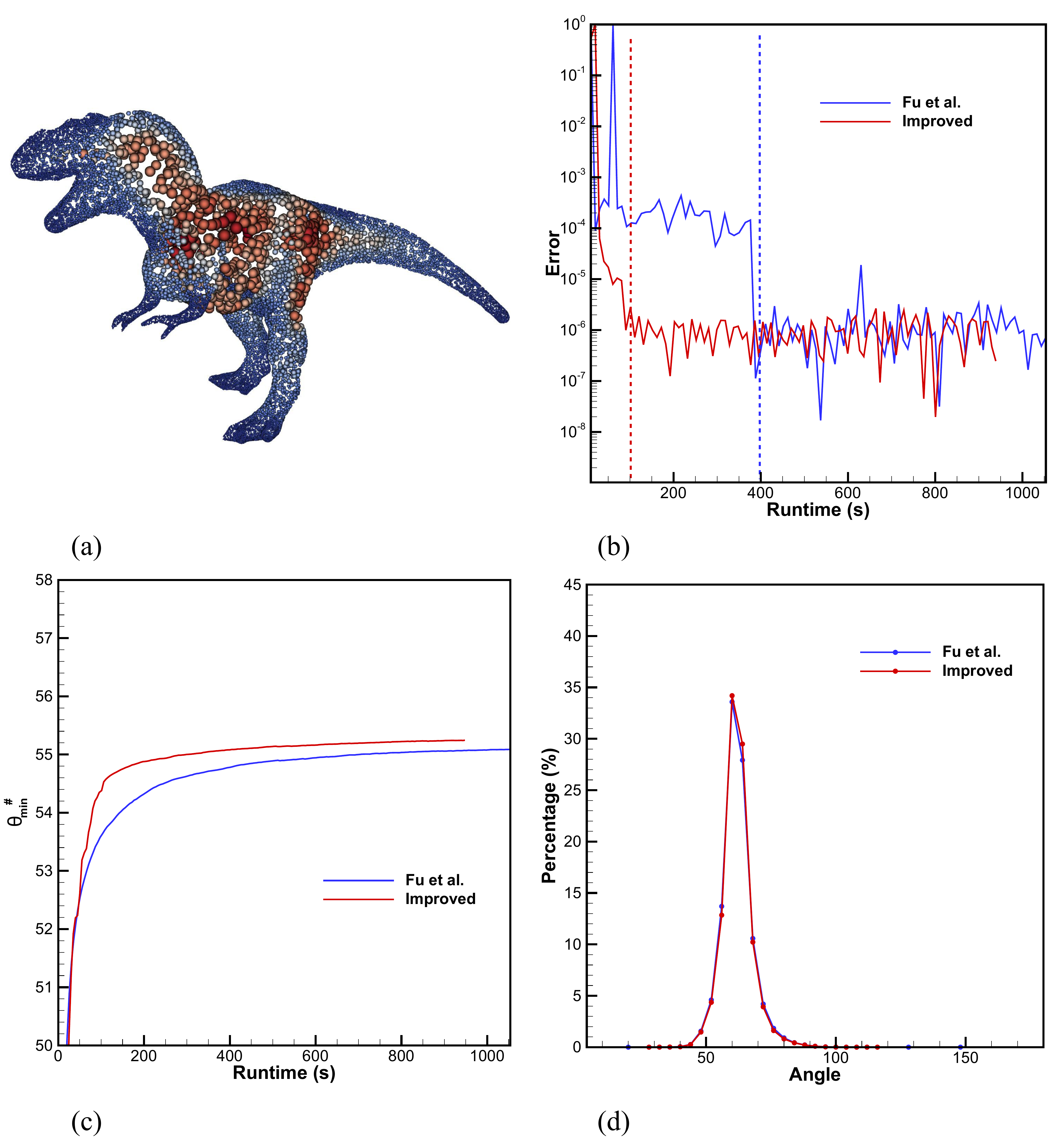}
		  \caption{T-Rex:(a) Initial particle distribution. The convergence histories of (b) $\overline{E}_{sys}$, (c) $\theta_{min}^{\#}$. (d) Histogram of the angle distribution.}
		\label{fig:Trex_stats}
		\end{figure}

		In order to evaluate the proposed method in more complex scenarios, we consider the geometry Tyrannosaurus rex, referred as ``T-Rex". The target-feature size function is defined on the geometry surface based on the smoothed curvature field. We set $h_{min}=0.1$ and $h_{max}=0.01$, which results in a total number of 21,243 particles. In this case, only surface meshes are generated, i.e. only $\mathbb{P}_s$ are considered in the simulation. The initial particle sampling is presented in Fig. \ref{fig:Trex_stats} (a) and colored with the target feature size.

\begin{figure}[H]
		  \centering
		    \includegraphics[width=1.0\textwidth]{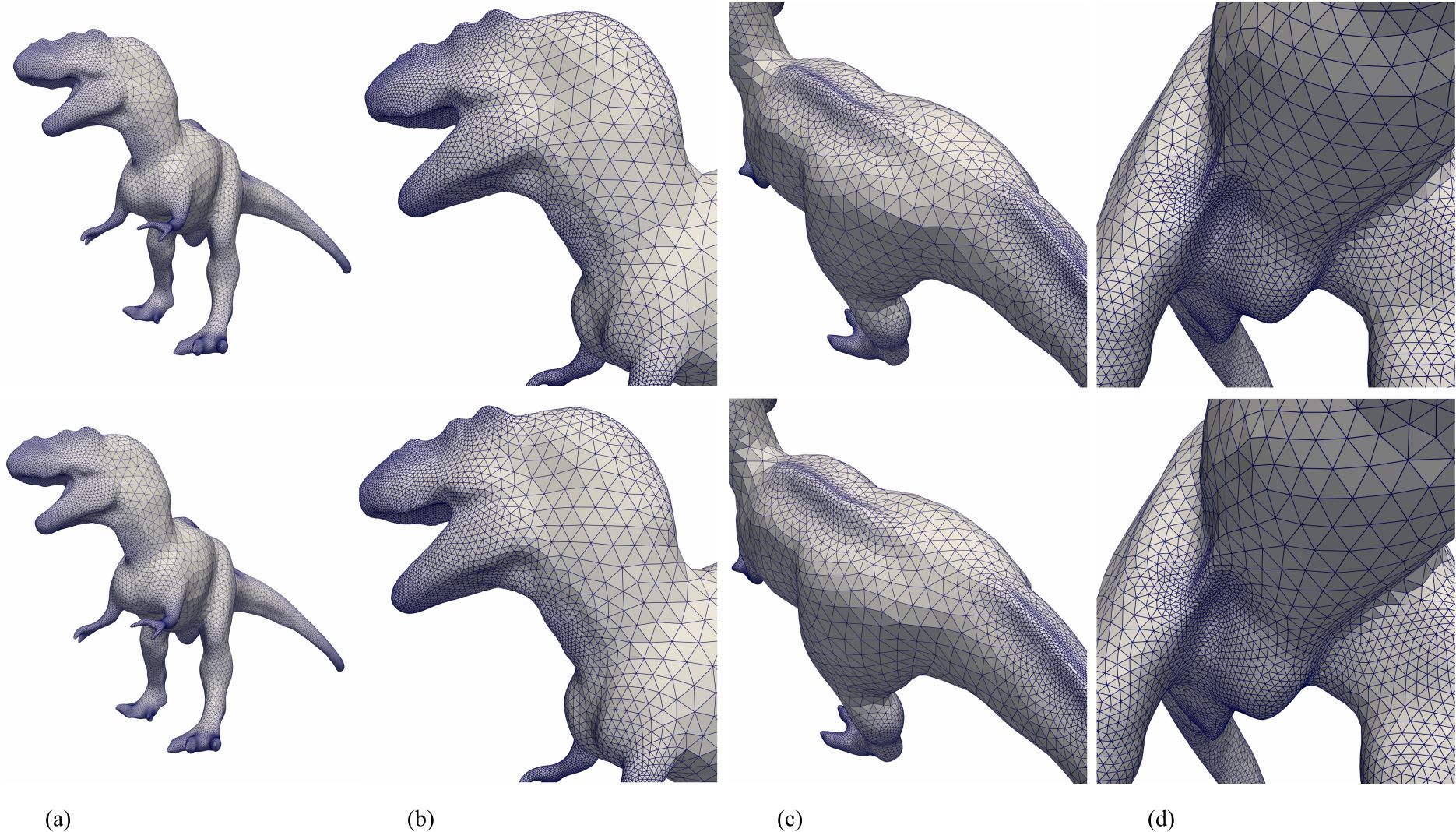}
		  \caption{T-Rex: (a) Delaunay triangulation of the final mesh and (b)(c)(d) the zoomed-in views at different camera positions. Upper row: results calculated by the algorithm developed by Fu et al. \cite{FU2019396}. Bottom row: results calculated by the proposed algorithm.}
		\label{fig:Trex_particle}
		\end{figure}

		The convergence histories of $\overline{E}_{sys}$ and $\theta_{min}^{\#}$ (Fig. \ref{fig:Trex_stats} (b) and (c)) suggest that both methods achieve the convergence successfully. The ``\textit{improved}" method achieves the target of Phase One after ca. 100s, while method \cite{FU2019396} takes ca. 400s (see Fig \ref{fig:Trex_stats} (b)), which results in a speedup factor of 4. From the triangulated results in Fig. \ref{fig:Trex_particle}, highly similar meshes are generated by both methods. A slightly better mesh quality of $G_{min}$, $\theta_{max}$ and $\theta_{min}$ are observed as shown in Table \ref{Tab:validation_tyra}.

		\begin{table}[h]
		\centering
		\caption{Mesh quality of the T-Rex case}
		\scriptsize
		\label{Tab:validation_tyra}
		\newcommand{\tabincell}[2]{\begin{tabular}{@{}#1@{}}#2\end{tabular}}
		\begin{tabular}{>{\centering\arraybackslash}m{1.2cm}
		                >{\centering\arraybackslash}m{0.6cm}
		                >{\centering\arraybackslash}m{0.6cm}
		                >{\centering\arraybackslash}m{1cm}
		                >{\centering\arraybackslash}m{1cm}
		                >{\centering\arraybackslash}m{1cm}
		                >{\centering\arraybackslash}m{1cm}
		                >{\centering\arraybackslash}m{1cm}
		                >{\centering\arraybackslash}m{1cm}
		                >{\centering\arraybackslash}m{1cm}}
		\hline
		 & $G_{avg}$ & $G_{min}$ & $\theta_{max}$ & $\theta_{min}$ & $\theta_{min}^{\#}$ & $\theta_{<30}$ & $N_{tri}$ & $runtime$ & $N_{iter}$\\ \hline
		 \textit{Fu et al.} & 0.94 &  0.35 &  133.91 & 22.52 & 54.87 & 5 & 42,429 & 478.43 & 8,900 \\
		 \textit{Improved}  & 0.94 &  0.48 &  117.88 & 23.90 & 54.83 & 6 & 42,443 & 176.96 & 3,500  \\
		\hline
		\end{tabular}
		\end{table}

	\subsection{Sphere}
	\label{S:validation_Sphere}

		In this section, the proposed method is extended to generate 3D tetrahedral meshes. A sphere with two target feature-size distributions are considered, which are referred as ``Sphere01" and ``Sphere02" respectively.

\begin{figure}[H]
		  \centering
		    \includegraphics[width=0.8\textwidth]{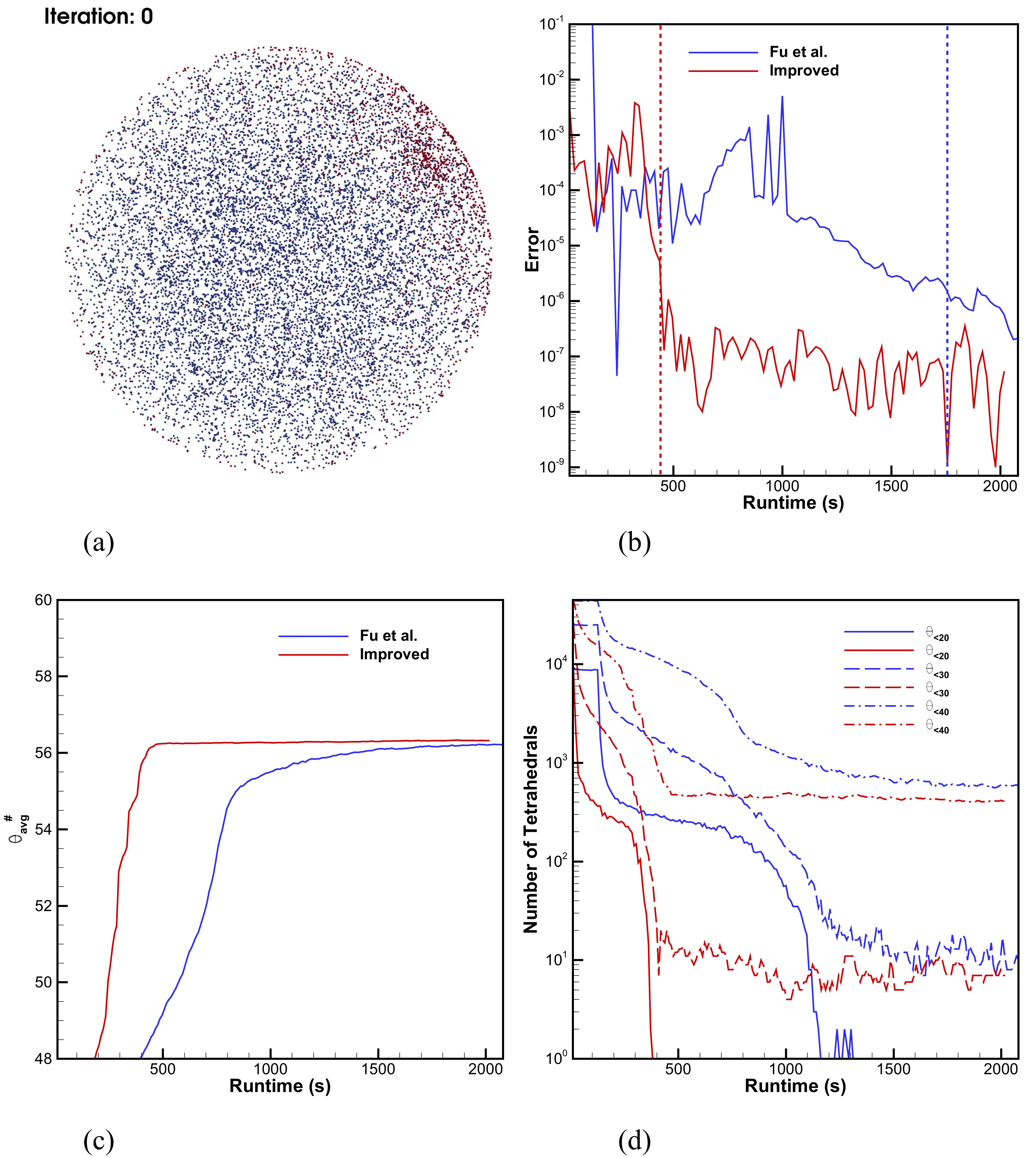}
		  \caption{Sphere01:(a) Initial particle distribution. The convergence histories of (b) $\overline{E}_{sys}$, (c) $\theta_{min}^{\#}$, and (d) $\theta_{<20}$, $\theta_{<30}$ and $\theta_{<40}$.}
		\label{fig:sphere01_stats_01}
		\end{figure}

		For ``Sphere01" case, we choose
		\begin{equation}
		\label{eq:h_t_sphere01}
			h_t(x,y)=\frac{h_{max}-h_{min}}{2R}\sqrt{(x-x_1)^2+(y-y_1)^2+(z-z_1)^2}+h_{min},
		\end{equation}
		where $R=45$ is the radius and $(x_1,y_1,z_1)$ is a point on the sphere surface. The minimum feature-size $h_{min}=0.375$ and the maximum feature-size $h_{max}=7.5$. The total number of particles is 12,540. The initial particle sampling is plotted in Fig. \ref{fig:sphere01_stats_01} (a).

\begin{figure}[H]
		  \centering
		    \includegraphics[width=0.8\textwidth]{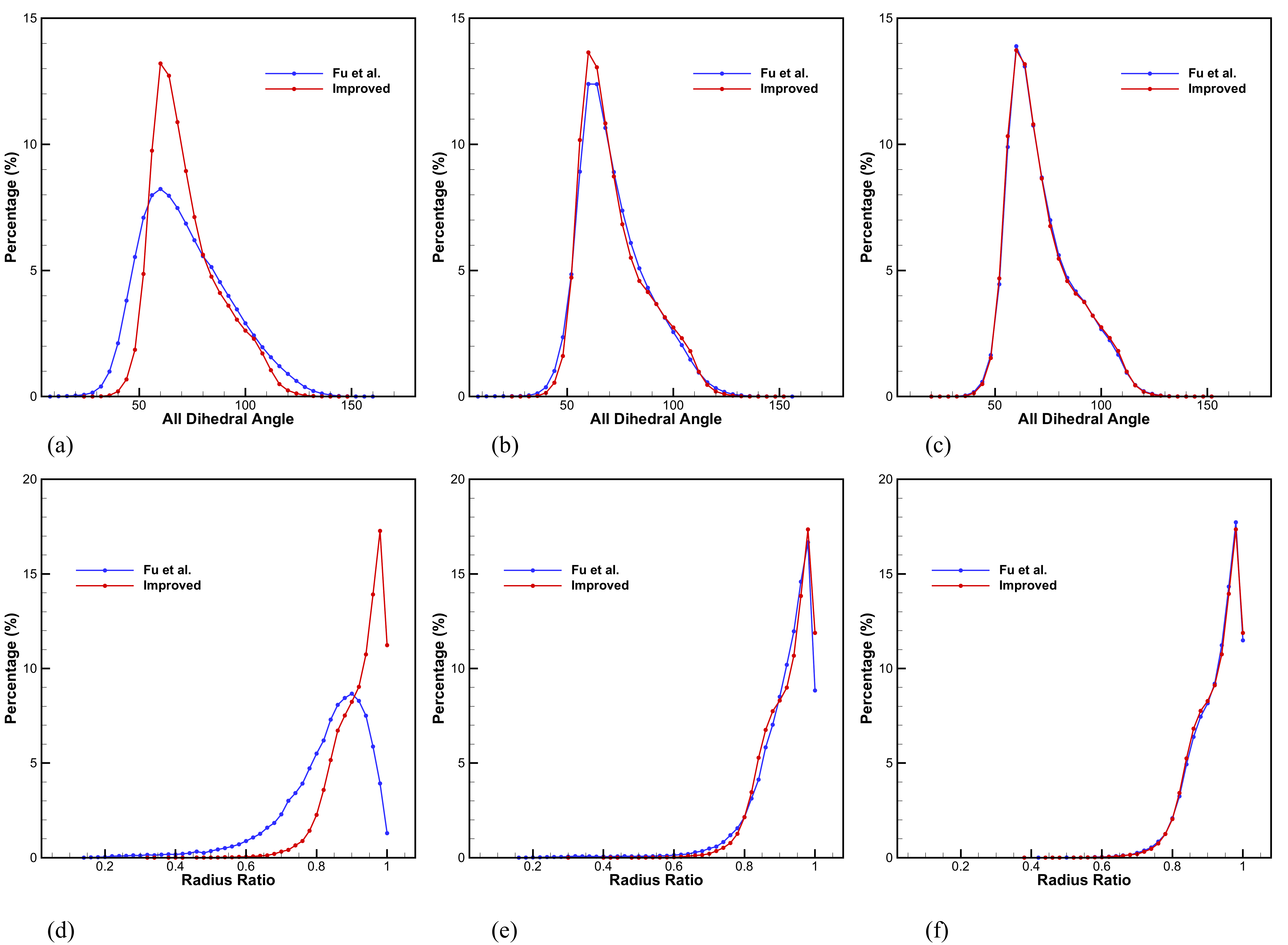}
		  \caption{Sphere01: Histogram of the dihedral angle distribution at (a) iteration 4000, (b) iteration 8000, and (c) iteration 20000. Histogram of the radius ratio distribution at (d) iteration 4000, (e) iteration 8000, and (f) iteration 20000.}
		\label{fig:sphere01_stats_02}
		\end{figure}

\begin{figure}[H]
		  \centering
		    \includegraphics[width=0.8\textwidth]{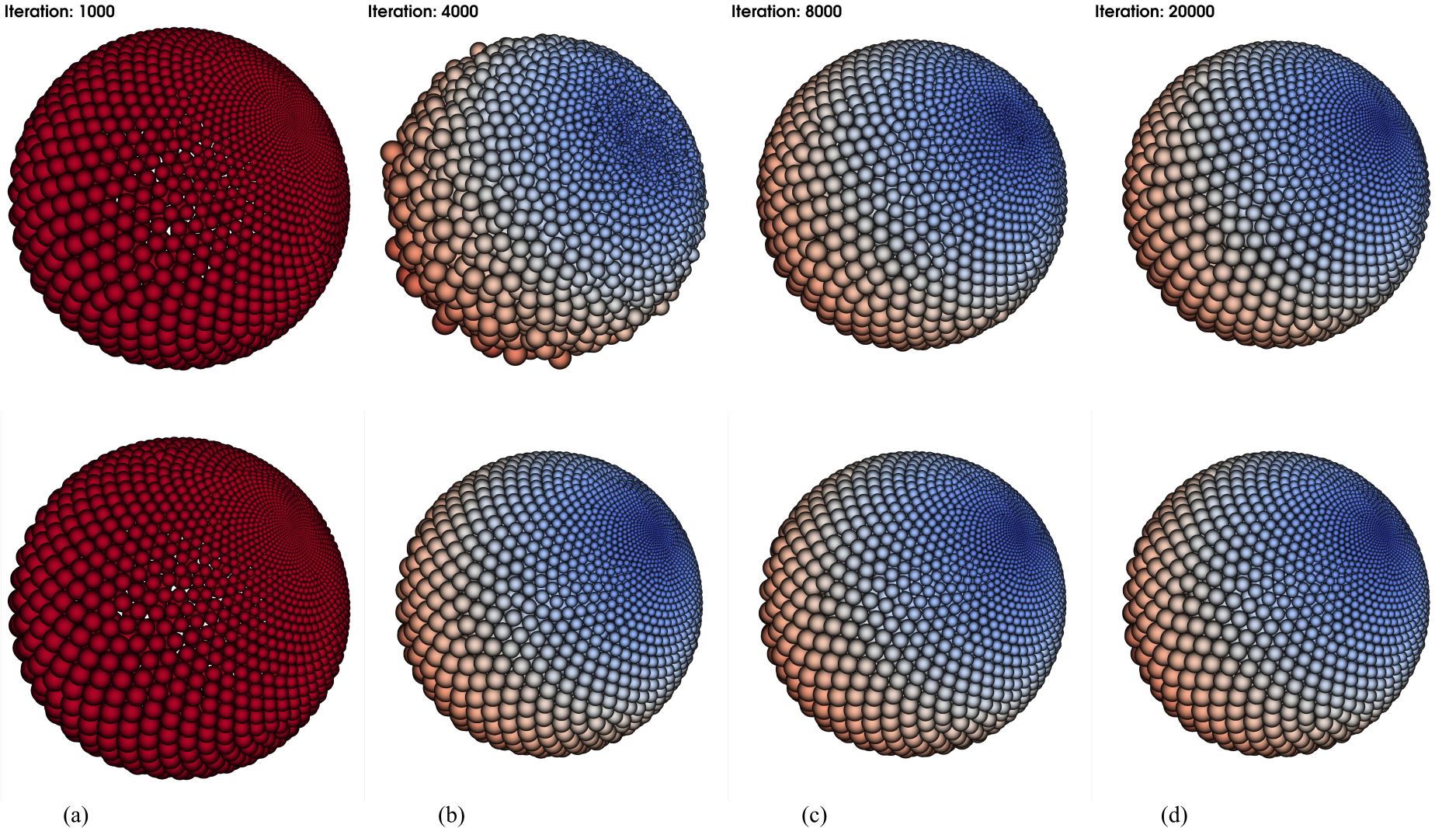}
		  \caption{Sphere01: Particle distributions at (a) iteration 1000, (b) iteration 4000, (c) iteration 8000 and (d) iteration 20000.  Upper row: results calculated by the algorithm developed by Fu et al. \cite{FU2019396}. Bottom row: results calculated by the proposed algorithm. Only particles belonging to positive cell ($\mathbb{C}_{+}$) are plotted in (b)(c)(d) and colored by target mesh-size.}
		\label{fig:sphere01_particle}
		\end{figure}

		From the time history of $\overline{E}_{sys}$ (see Fig. \ref{fig:sphere01_stats_01} (b)), the convergence of Phase One is achieved after ca. 430s for the ``\textit{improved}" method and 1750s for method \cite{FU2019396}. The ``\textit{improved}" method reaches the optimized mesh quality remarkably faster than method \cite{FU2019396}, which is revealed from the convergence histories of $\theta_{min}^{\#}$ (Fig. \ref{fig:sphere01_stats_01} (c)) and $\theta_{<20}$/$\theta_{<30}$/$\theta_{<40}$. (Fig. \ref{fig:sphere01_stats_01} (d)). The histograms of dihedral angle and radius ratio distribution (Fig. \ref{fig:sphere01_stats_02}) show that the underlying measured statistics remain almost constant after 4000 iterations for the ``\textit{improved}" method. At 20000 iterations, both methods achieve approximately the same distributions. From the snapshots of the simulation at different instants (see Fig. \ref{fig:sphere01_particle}), the same conclusions can be made. Except for the improvement achieved in runtime, the ``\textit{improved}" method also exhibits slightly better mesh quality, which can be seen from Table \ref{Tab:validation_Sphere01}.

		\begin{table}[h]
		\centering
		\caption{Mesh quality of the Sphere01 case}
		\scriptsize
		\label{Tab:validation_Sphere01}
		\newcommand{\tabincell}[2]{\begin{tabular}{@{}#1@{}}#2\end{tabular}}
		\begin{tabular}{>{\centering\arraybackslash}m{1.2cm}
		                >{\centering\arraybackslash}m{1.4cm}
		                >{\centering\arraybackslash}m{1.4cm}
		                >{\centering\arraybackslash}m{0.6cm}
		                >{\centering\arraybackslash}m{0.6cm}
		                >{\centering\arraybackslash}m{0.6cm}
		                >{\centering\arraybackslash}m{0.6cm}
		                >{\centering\arraybackslash}m{0.6cm}
		                >{\centering\arraybackslash}m{0.6cm}
		                >{\centering\arraybackslash}m{0.8cm}
		                >{\centering\arraybackslash}m{0.8cm}}
		\hline
		 & $\theta_{min}/\theta_{max}$ & $\gamma_{min}/\gamma_{avg}$ & $\theta_{min}^{\#}$ & $\theta_{<10}$ & $\theta_{<20}$ & $\theta_{<30}$ & $\theta_{<40}$ & $N_{tet}$ & $runtime$ & $N_{iter}$\\ \hline
		 \textit{Fu et al.} & 22.44/152.11 &  0.24/0.91 & 56.17 & 0 & 0 & 13 & 592 & 68,724 & 1,782.80 & 17,000 \\
		 \textit{Improved}  & 22.33/150.78 &  0.32/0.91 & 56.24 & 0 & 0 & 12 & 478 & 68,757 &  448.35 & 4,600 \\
		\hline
		\end{tabular}
		\end{table}

		\begin{figure}[H]
		  \centering
		    \includegraphics[width=0.8\textwidth]{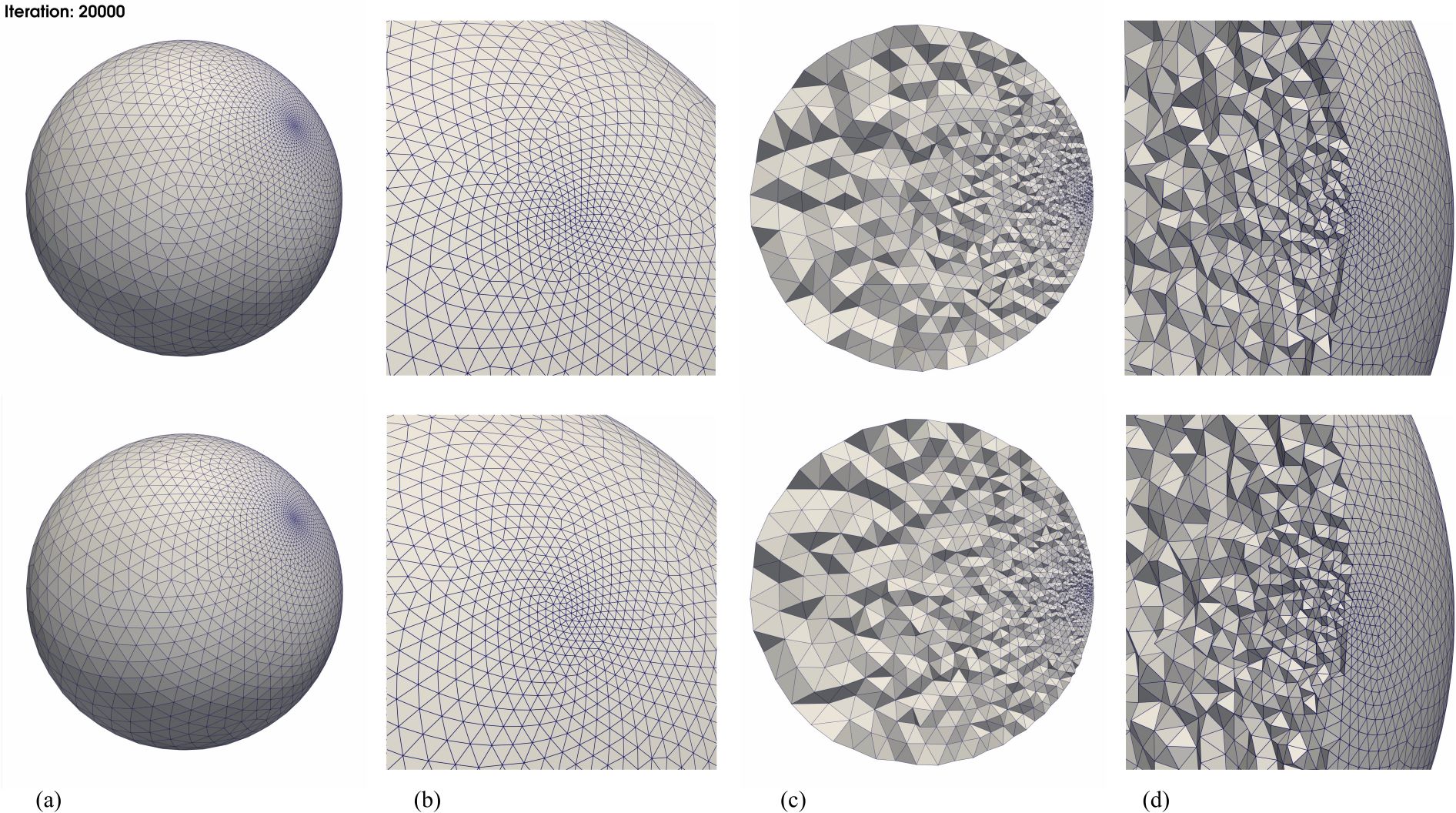}
		  \caption{Sphere01: Tetrahedralization of the final mesh. (a) The full-size view. (b) The zoomed-in view. (c)(d) Clipped views. Upper row: results calculated by the algorithm developed by Fu et al. \cite{FU2019396}. Bottom row: results calculated by the proposed algorithm.}
		  \label{fig:sphere01_mesh}
		\end{figure}

		For the case ``Sphere02", we follow \cite{ni2017sliver}. The target feature-size function is defined as 
		\begin{equation}
		\label{eq:h_t_sphere01}
			h_t(x,y)=0.025+0.2\mid\sqrt{x^2+y^2+z^2}-0.5\mid.
		\end{equation}
		The minimum feature-size $h_{min}=0.025$ and the maximum feature-size $h_{max}=0.125$. The result is compared with three methods in \cite{ni2017sliver}, i.e. the Gradient-based Shape Matching (denoted as ``\textit{GSM}"), the GSM coupled with a particle-based method \cite{zhong2013particle} (denoted as ``\textit{particle+GSM}") and the particle-based method (denoted as ``\textit{particle}").

\begin{figure}[H]
		  \centering
		    \includegraphics[width=0.8\textwidth]{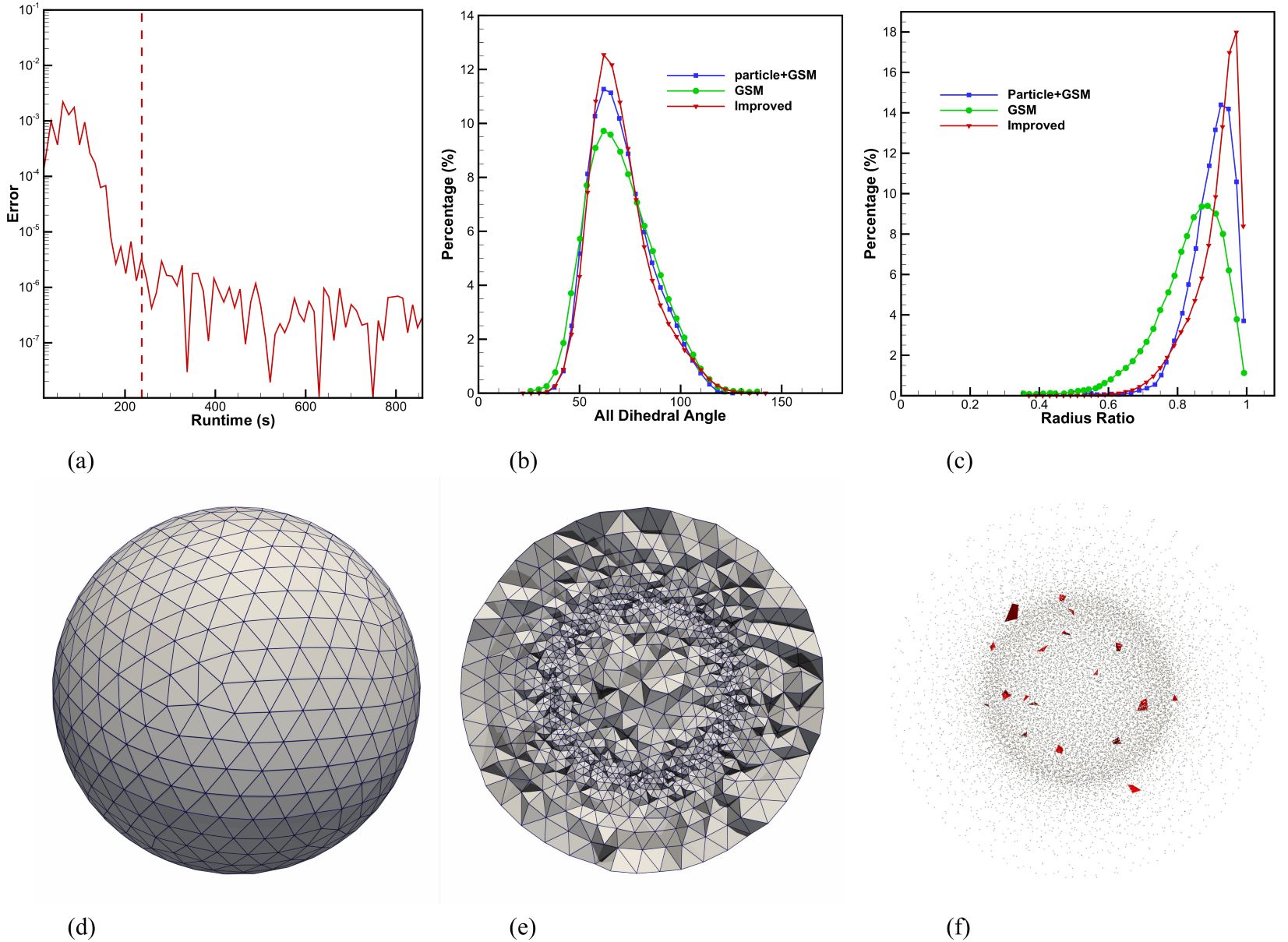}
		  \caption{Sphere02: (a) The convergence histories of $\overline{E}_{sys}$. Comparison of (b) dihedral angle distribution histogram and (c) radius ratio distribution histogram between the results in \cite{ni2017sliver} and result obtained by the proposed algorithm. (d) Tetrahedralization of the final mesh. (e) Clipped view of the final mesh. (f) Tetrahedrons with dihedral angle below $30^{\circ}$.}
		\label{fig:Sphere02_compare}
		\end{figure}

		Fig. \ref{fig:Sphere02_compare} (a) illustrates the time history of $\overline{E}_{sys}$. The target of Phase One achieves after ca. 240s. The dihedral angle and the radius ratio distribution histograms from the ``\textit{improved}" method, the ``\textit{GSM}" method and the ``\textit{particle+GSM}" method are plotted together in Fig. \ref{fig:Sphere02_compare} (b) and (c) respectively. It can be observed that the ``\textit{improved}" method generates the largest proportion of regular tetrahedrons than the other two methods. For the number of ``slivers" generated, the ``\textit{improved}" method outperforms the ``\textit{particle}" method and the ``\textit{GSM}" method (see Table \ref{Tab:validation_Sphere02}). The ``\textit{particle+GSM}" presents the best mesh quality overall. However, in terms of computational cost, the ``\textit{improved}" method only takes 236.42s to achieve the mesh quality, while 8,787.22s are required for the ``\textit{particle+GSM}" method (see Table \ref{Tab:validation_Sphere02}). With slightly more computational effort, the proposed method achieves a speedup factor of 37.

		\begin{table}[h]
		\centering
		\caption{Mesh quality of the Sphere02 case}
		\scriptsize
		\label{Tab:validation_Sphere02}
		\newcommand{\tabincell}[2]{\begin{tabular}{@{}#1@{}}#2\end{tabular}}
		\begin{tabular}{>{\centering\arraybackslash}m{1.7cm}
		                >{\centering\arraybackslash}m{1.5cm}
		                >{\centering\arraybackslash}m{1.4cm}
		                >{\centering\arraybackslash}m{0.6cm}
		                >{\centering\arraybackslash}m{0.6cm}
		                >{\centering\arraybackslash}m{0.6cm}
		                >{\centering\arraybackslash}m{0.6cm}
		                >{\centering\arraybackslash}m{0.8cm}
		                >{\centering\arraybackslash}m{1.0cm}
		                >{\centering\arraybackslash}m{0.6cm}}
		\hline
		 & $\theta_{min}/\theta_{max}$ & $\gamma_{min}/\gamma_{avg}$ & $\theta_{<10}$ & $\theta_{<20}$ & $\theta_{<30}$ & $\theta_{<40}$ & $N_{tet}$ & $runtime$\footnotemark & $N_{iter}$\\ \hline
		 \textit{Improved}     & 26.25/143.78 &  0.32/0.91    & 0   & 0     & 18    & 1,873  & 110,706 &  236.42   & 3800 \\
		 \textit{GSM}          & 21.1/138     &  0.355/0.837  & 0   & 0     & 118   & 5,887  & 103,485 &  7,840.78 & --- \\
		 \textit{particle+GSM} & 30.2/130     &  0.531/0.900  & 0   & 0     & 0     & 1,092  & 106,106 &  8,787.22 & --- \\
		 \textit{particle}     & 1.03/178     &  0.0203/0.856 & 372 & 1,718 & 4,655 & 12,282 & 111,505 &  410.37   & --- \\
		\hline
		\end{tabular}
		\end{table}
		\footnotetext[5]{The runtime of the ``\textit{GSM}", ``\textit{particle+GSM}" and ``\textit{particle}" method are measured by an Intel\textsuperscript{\textregistered} Xeon\textsuperscript{\textregistered} CPU(E5645, 2.40GHz, 12 threads).}

	\subsection{Cube}
	\label{S:validation_cube}

\begin{figure}[H]
		  \centering
		    \includegraphics[width=0.8\textwidth]{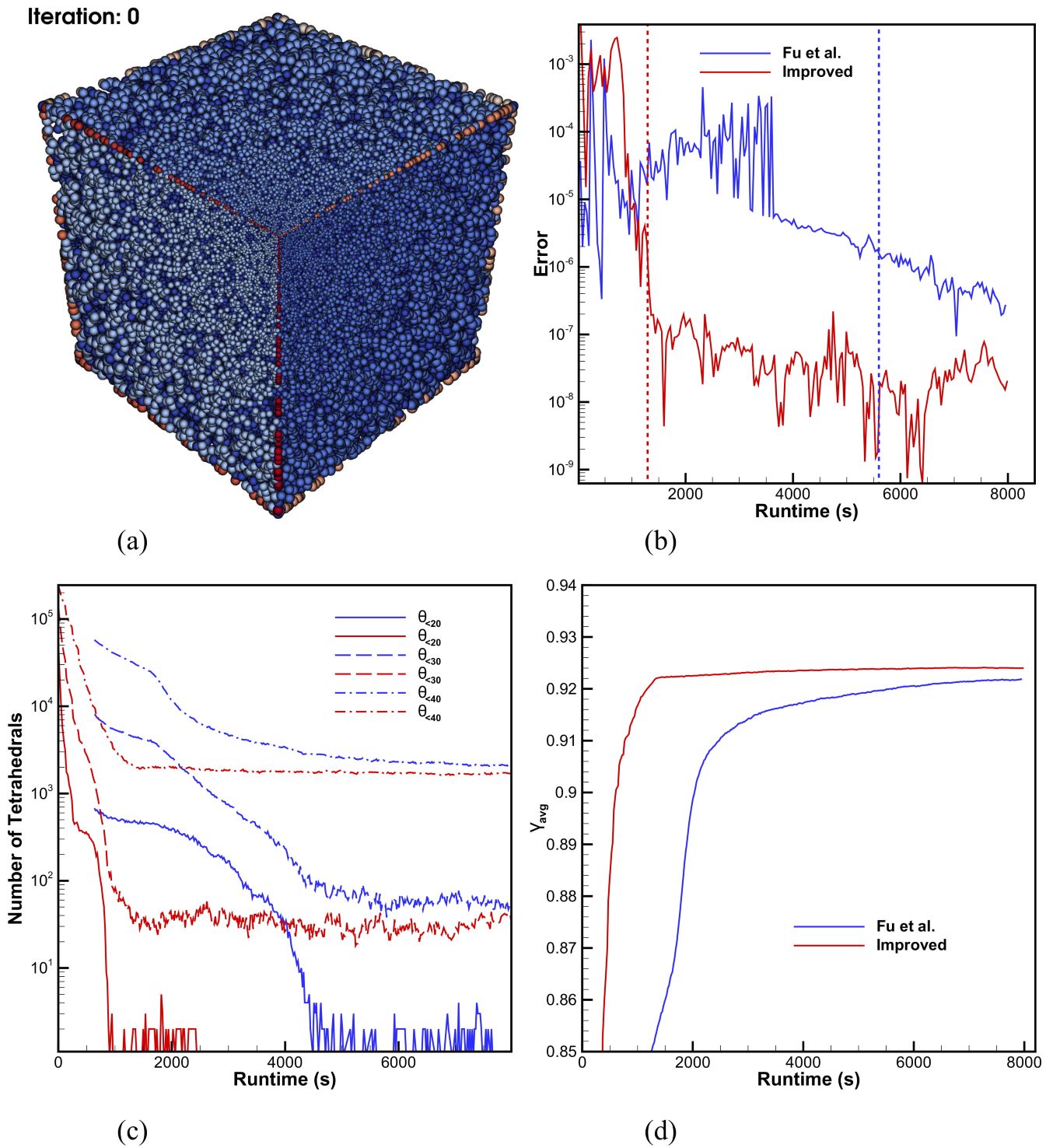}
		  \caption{Cube:(a) Initial particle distribution. The convergence histories of (b) $\overline{E}_{sys}$, (c) $\theta_{<20}$, $\theta_{<30}$ and $\theta_{<40}$, and (d) $\gamma_{avg}$.}
		\label{fig:cube_stats_01}
		\end{figure}

		Next, we consider a cube. In this case, multiple sharp edges and singularity points are presented. The target feature-size function is defined as 
		\begin{equation}
		\label{eq:h_t_cube}
			h_t(x,y)=\frac{h_{max}-h_{min}}{100\sqrt{3}}\sqrt{(x-100)^2+(y-100)^2+(z-100)^2}+h_{min},
		\end{equation}
		where $h_{min}=0.488$, $h_{max}=4.88$, and $(x,y,z)\in[0,100]$. The resulting number of particle is 66,038. The initial particle distribution is plotted in Fig. \ref{fig:cube_stats_01} (a).

\begin{figure}[H]
		  \centering
		    \includegraphics[width=0.8\textwidth]{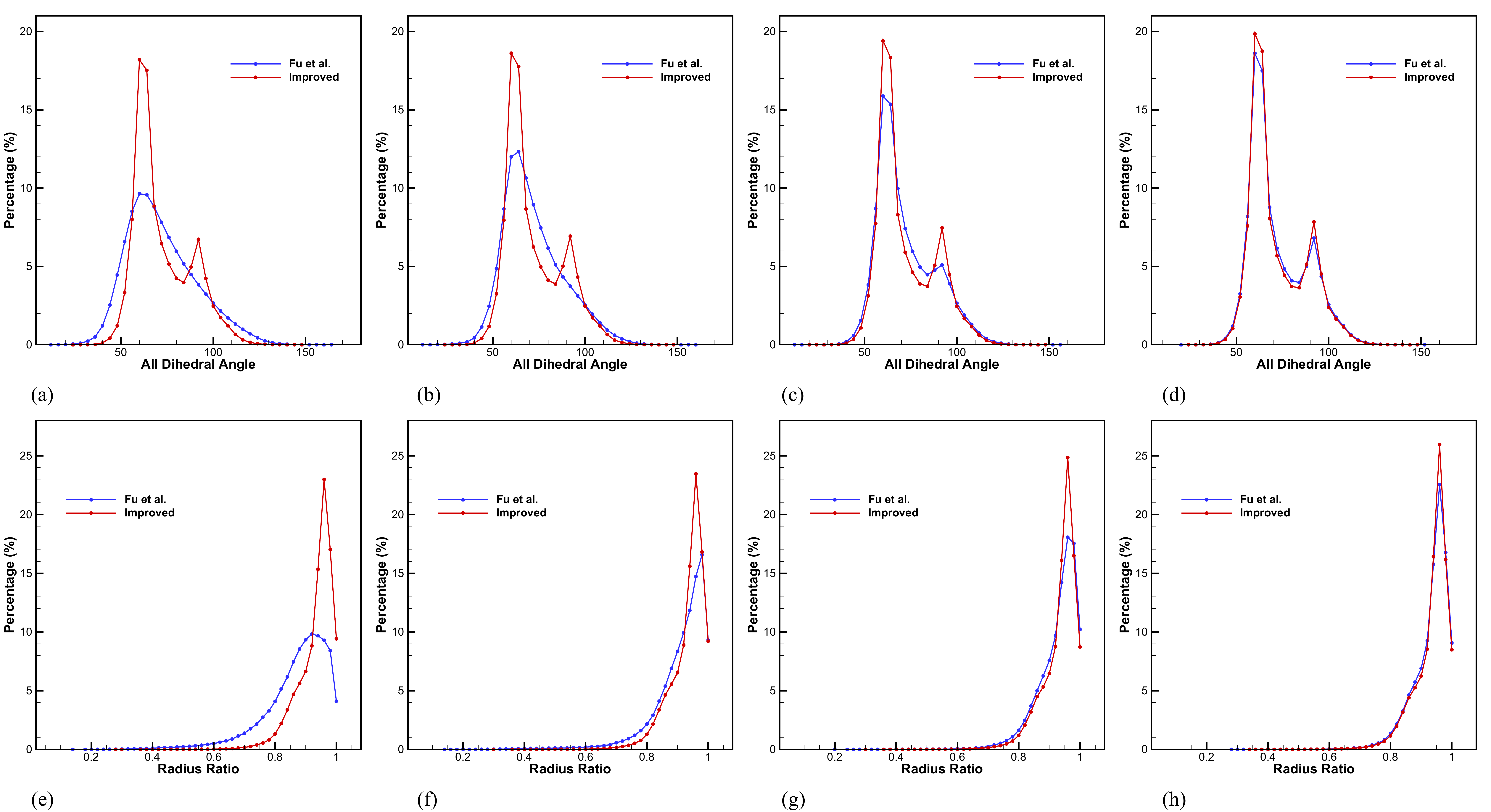}
		  \caption{Cube: Histogram of the dihedral angle distribution at (a) iteration 7000, (b) iteration 10000, (c) iteration 20000, and (d) iteration 40000. Histogram of the radius ratio distribution at (e) iteration 7000, (f) iteration 10000, (g) iteration 20000, and (h) iteration 40000.}
		\label{fig:cube_stats_02}
		\end{figure}

		The results and comparisons are presented in Table \ref{Tab:validation_Cube}, Fig. \ref{fig:cube_stats_01} and Fig. \ref{fig:cube_stats_02}. The conclusions are consistent with previous observations. A speedup factor of 4.3 (1,300s v.s. 5,600s) is obtained comparing the wall-clock time used for achieving the convergence. In terms of mesh quality, the ``\textit{improved}" method outperforms method \cite{FU2019396} slightly in each category.

		\begin{table}[h]
		\centering
		\caption{Mesh quality of the Cube case}
		\scriptsize
		\label{Tab:validation_Cube}
		\newcommand{\tabincell}[2]{\begin{tabular}{@{}#1@{}}#2\end{tabular}}
		\begin{tabular}{>{\centering\arraybackslash}m{1.2cm}
		                >{\centering\arraybackslash}m{1.4cm}
		                >{\centering\arraybackslash}m{1.4cm}
		                >{\centering\arraybackslash}m{0.5cm}
		                >{\centering\arraybackslash}m{0.5cm}
		                >{\centering\arraybackslash}m{0.5cm}
		                >{\centering\arraybackslash}m{0.5cm}
		                >{\centering\arraybackslash}m{0.6cm}
		                >{\centering\arraybackslash}m{0.8cm}
		                >{\centering\arraybackslash}m{0.8cm}
		                >{\centering\arraybackslash}m{0.8cm}}
		\hline
		 & $\theta_{min}/\theta_{max}$ & $\gamma_{min}/\gamma_{avg}$ & $\theta_{min}^{\#}$ & $\theta_{<10}$ & $\theta_{<20}$ & $\theta_{<30}$ & $\theta_{<40}$ & $N_{tet}$ & $runtime$ & $N_{iter}$\\ \hline
		 \textit{Fu et al.} & 20.88/151.92 &  0.24/0.91 & 56.63 & 0 & 0 & 51 & 2,421 & 367,908 & 5,729.70 & 28,800 \\
		 \textit{Improved}  & 23.07/146.74 &  0.27/0.92 & 56.88 & 0 & 0 & 28 & 1,913 & 367,839 & 1,437.30 & 7,000 \\
		\hline
		\end{tabular}
		\end{table}

\begin{figure}[H]
		  \centering
		    \includegraphics[width=0.8\textwidth]{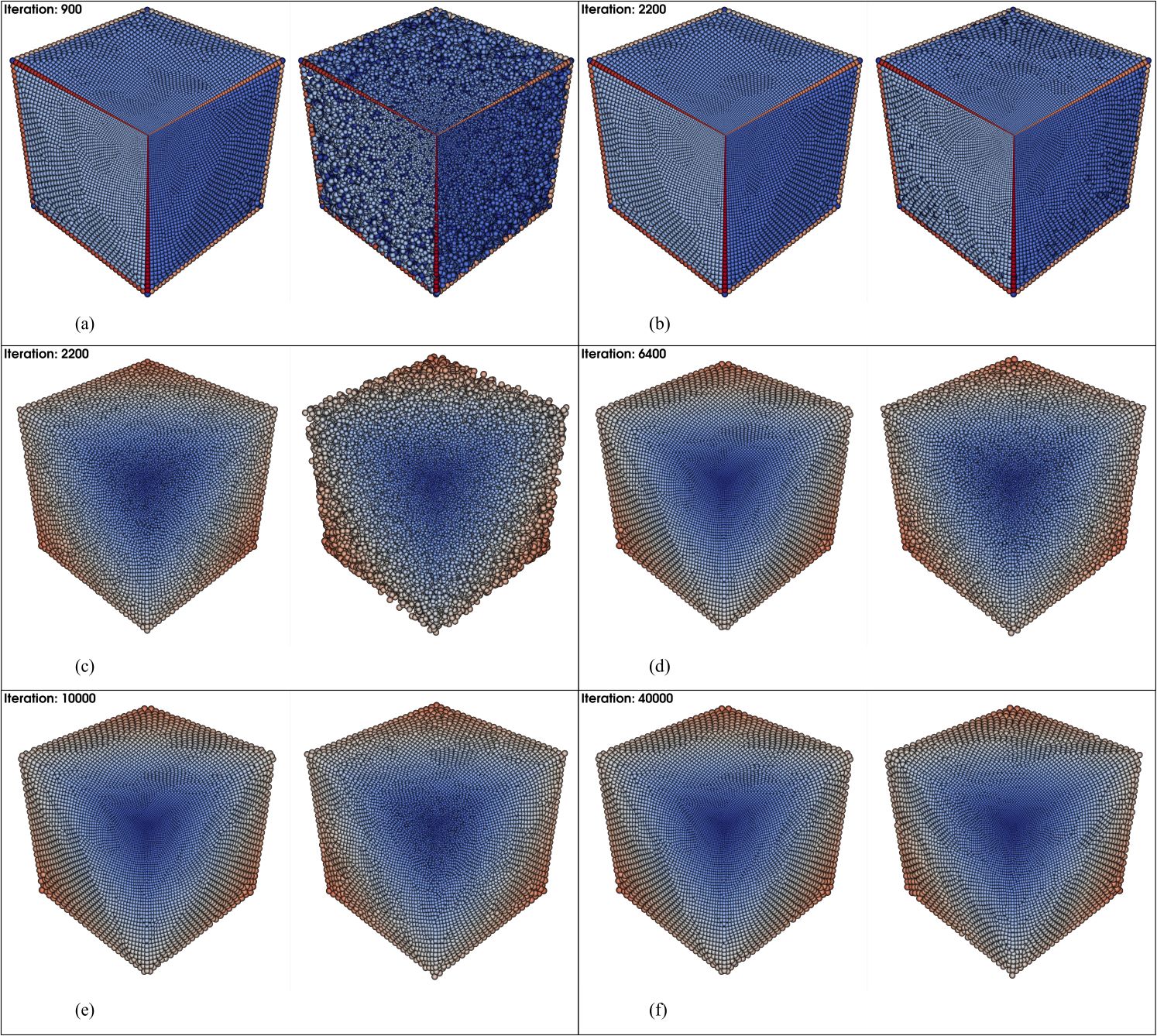}
		  \caption{Cube: Particle distributions at (a) iteration 900, (b)(c) iteration 2200, (d) iteration 6400, (e) iteration 10000, and (f) iteration 40000.  In each sub-figure, left: result calculated by the proposed algorithm; right: result calculated by the algorithm developed by Fu et al. \cite{FU2019396}. Only particles belonging to positive cell ($\mathbb{C}_{+}$) are plotted in (c)(d)(e)(f) and colored by target mesh-size.}
		\label{fig:cube_particle}
		\end{figure}

		The convergence curve of $\overline{E}_{sys}$ in Fig. \ref{fig:cube_stats_01} (a) reveals that, benefiting from the correction term and the accumulation of particle momentum, the ``\textit{improved}" method converges significantly faster than method \cite{FU2019396}. From the snapshots of particle distributions at various instants of the simulation (see Fig. \ref{fig:cube_particle}), we can see that at iteration No. 900, the equilibrium state of all $\mathbb{P}_s$ and $\mathbb{P}_c$ has been achieved for the ``\textit{improved}" method. For method \cite{FU2019396}, only convergence of $\mathbb{P}_c$ is achieved, since the particles are evolved following a dimensional sequence. The same observation can be made for iteration No. 2200, where $\mathbb{P}_s$ converge to the optimized positions while $\mathbb{P}_+$ are remained in the original positions. Conversely, for the ``\textit{improved}" method, particles of all features are evolved together. This can be observed from Fig. \ref{fig:cube_particle} (b) and (c). Consequently a faster convergence is achieved.

		\begin{figure}[H]
		  \centering
		    \includegraphics[width=0.8\textwidth]{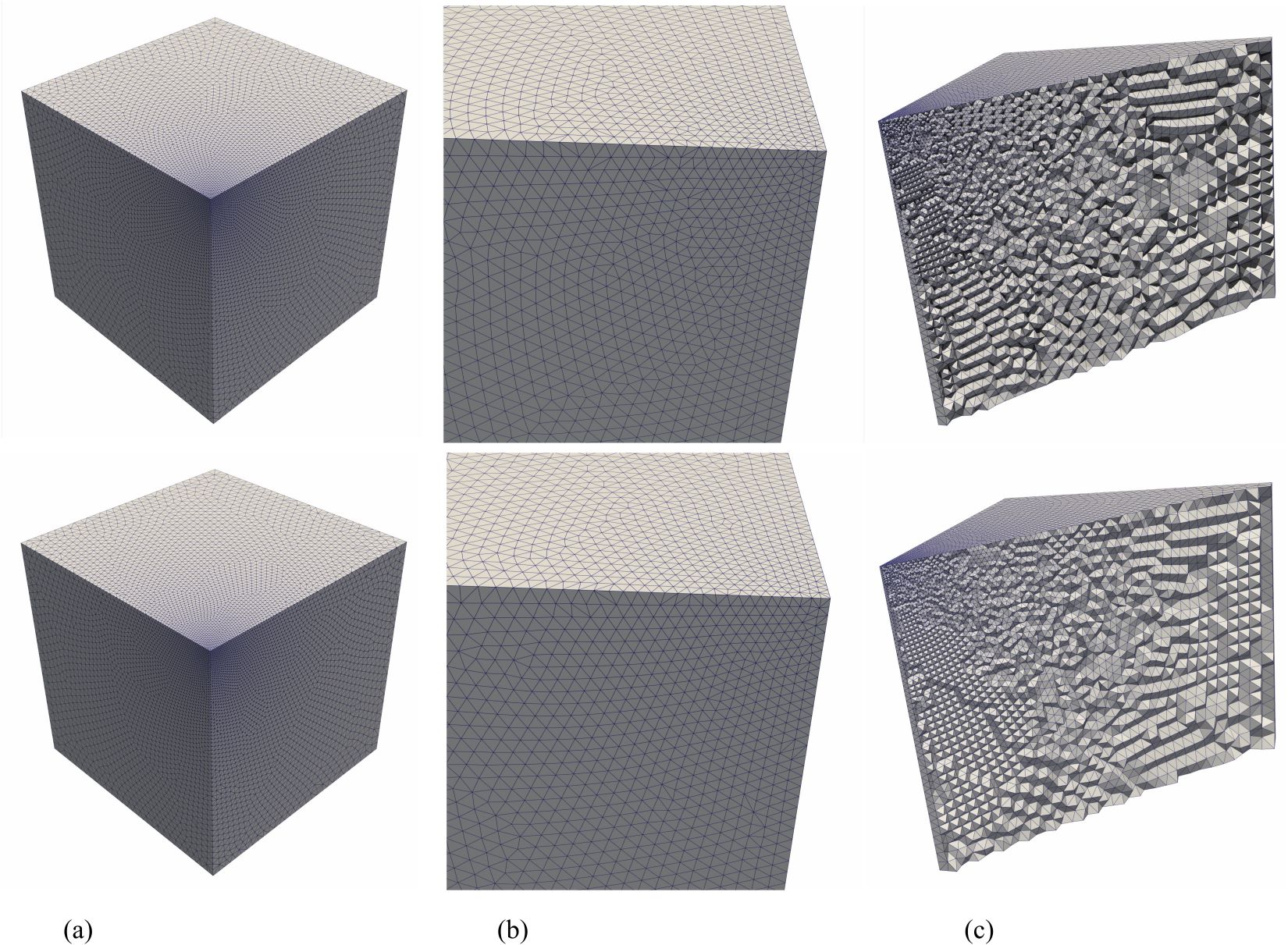}
		  \caption{Cube: Tetrahedralization of the final mesh. (a) The full-size view. (b) The zoomed-in view. (c) Clipped view. Upper row: results calculated by the algorithm developed by Fu et al. \cite{FU2019396}. Bottom row: results calculated by the proposed algorithm.}
		\label{fig:cube_mesh}
		\end{figure}

	\subsection{Spur gear}
	\label{S:validation_gear}

		Lastly a complex geometry of spur gear developed for gear lubrication tests \cite{hoehn2008test} is considered. The size of computational domain is $[86.4\times17.6\times86.4]$. A total number of 133 features are introduced in the model. The same target feature-size function is defined following \cite{ji2019consistent}. Two situations are simulated. In ``Gear01", only surface meshes are constructed, while in ``Gear02", both triangular and tetrahedral meshes are generated. The maximum and minimum feature-size for both cases are set the same as $h_{min}=0.4$ and $h_{max}=3.2$. The resulting total number of particles are 42,647 and 169,096 respectively. The objective of this case is to test the robustness of the proposed feature-aware SPH in the mesh generation of complex geometries.

\begin{figure}[H]
		  \centering
		    \includegraphics[width=0.8\textwidth]{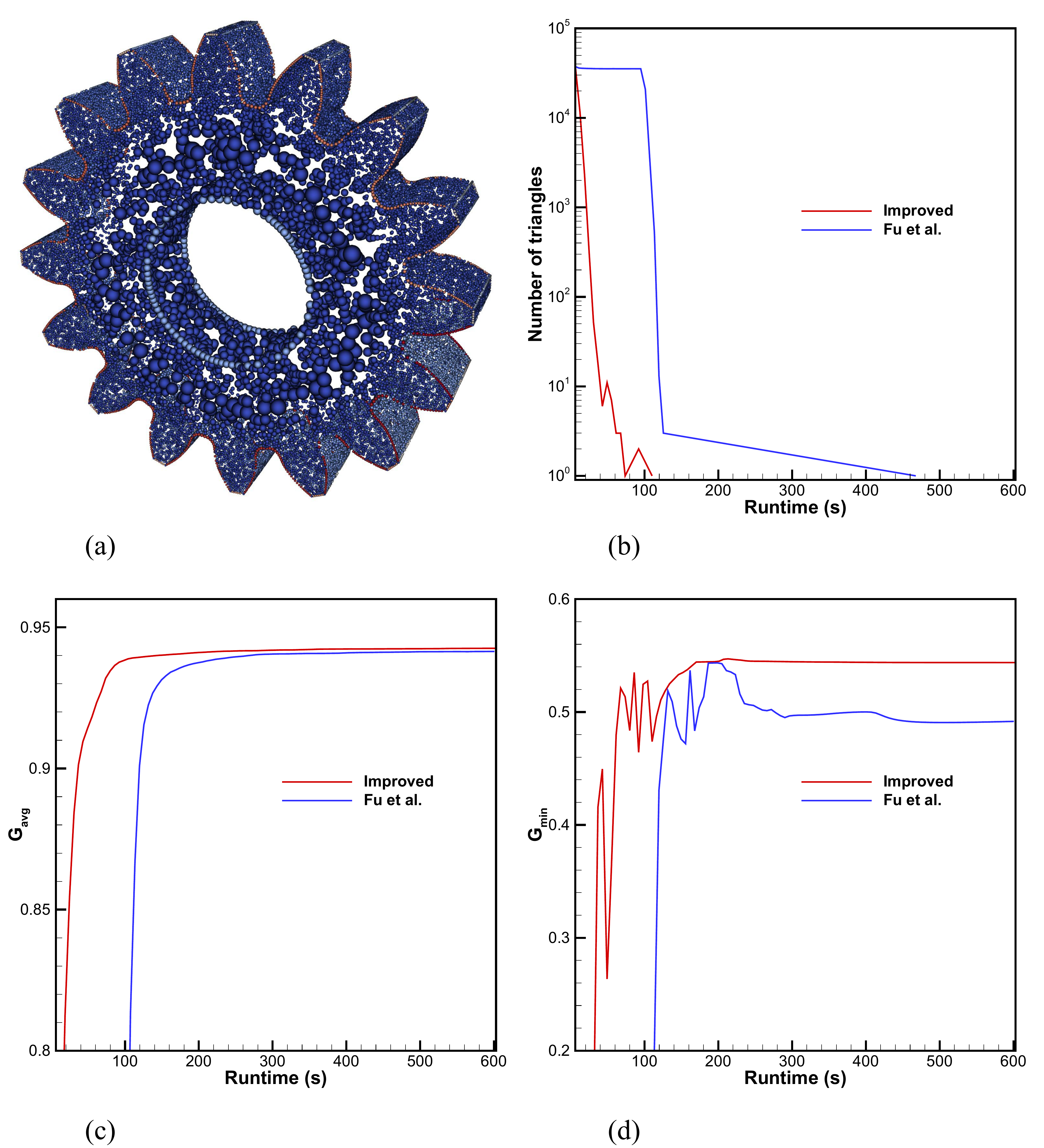}
		  \caption{Gear01:(a) Initial particle distribution. The convergence histories of (b) $\theta_{<30}$, (c) $G_{avg}$, and (d) $G_{min}$.}
		\label{fig:gear_stats}
		\end{figure}

\begin{figure}[H]
		  \centering
		    \includegraphics[width=0.8\textwidth]{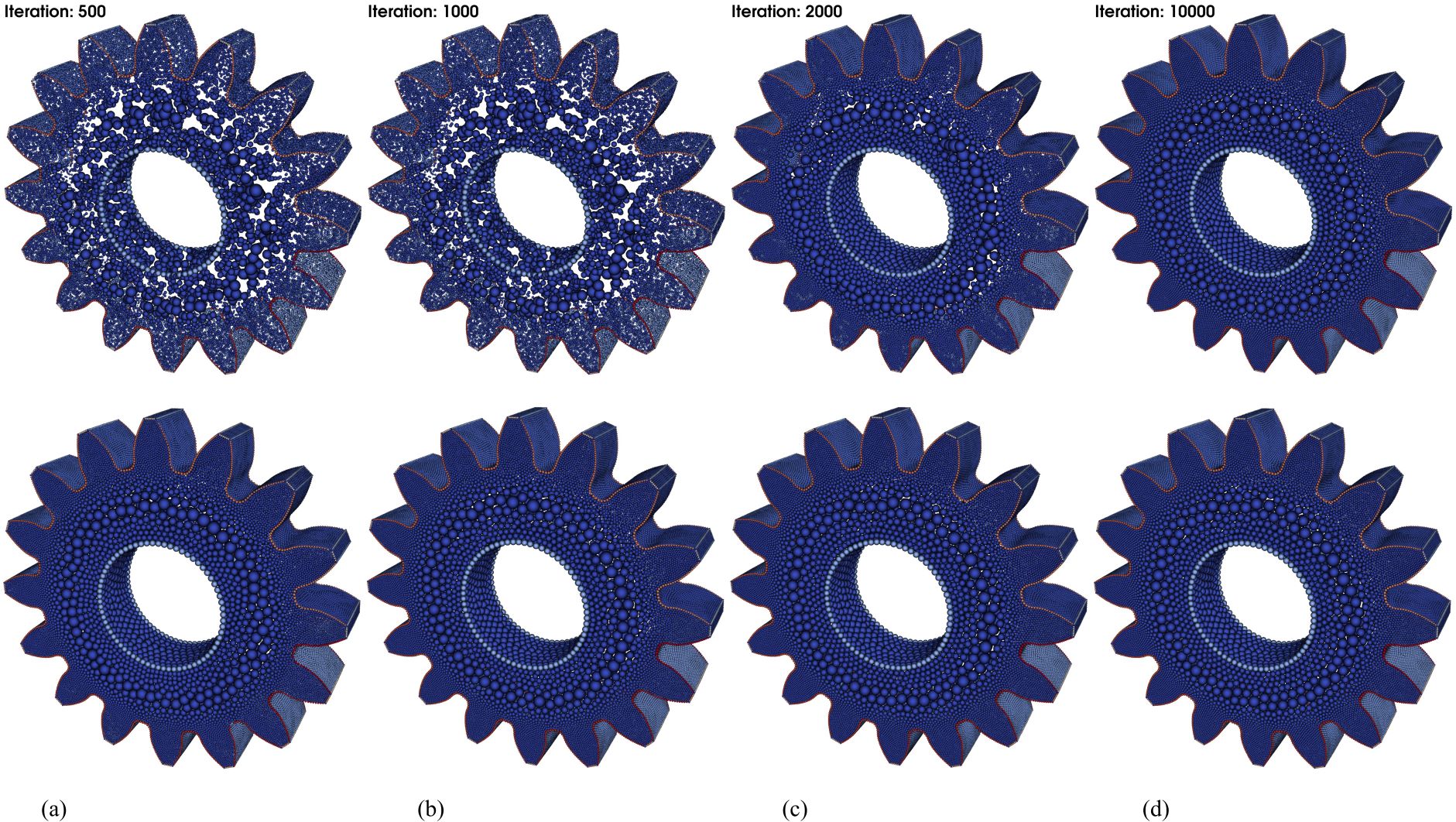}
		  \caption{Gear01: Particle distributions at (a) iteration 500, (b) iteration 1000, (c) iteration 2000 and (d) iteration 10000.  Upper row: results calculated by the algorithm developed by Fu et al. \cite{FU2019396}. Bottom row: results calculated by the proposed algorithm.}
		\label{fig:gear_particle}
		\end{figure}

\begin{figure}[H]
		  \centering
		    \includegraphics[width=0.8\textwidth]{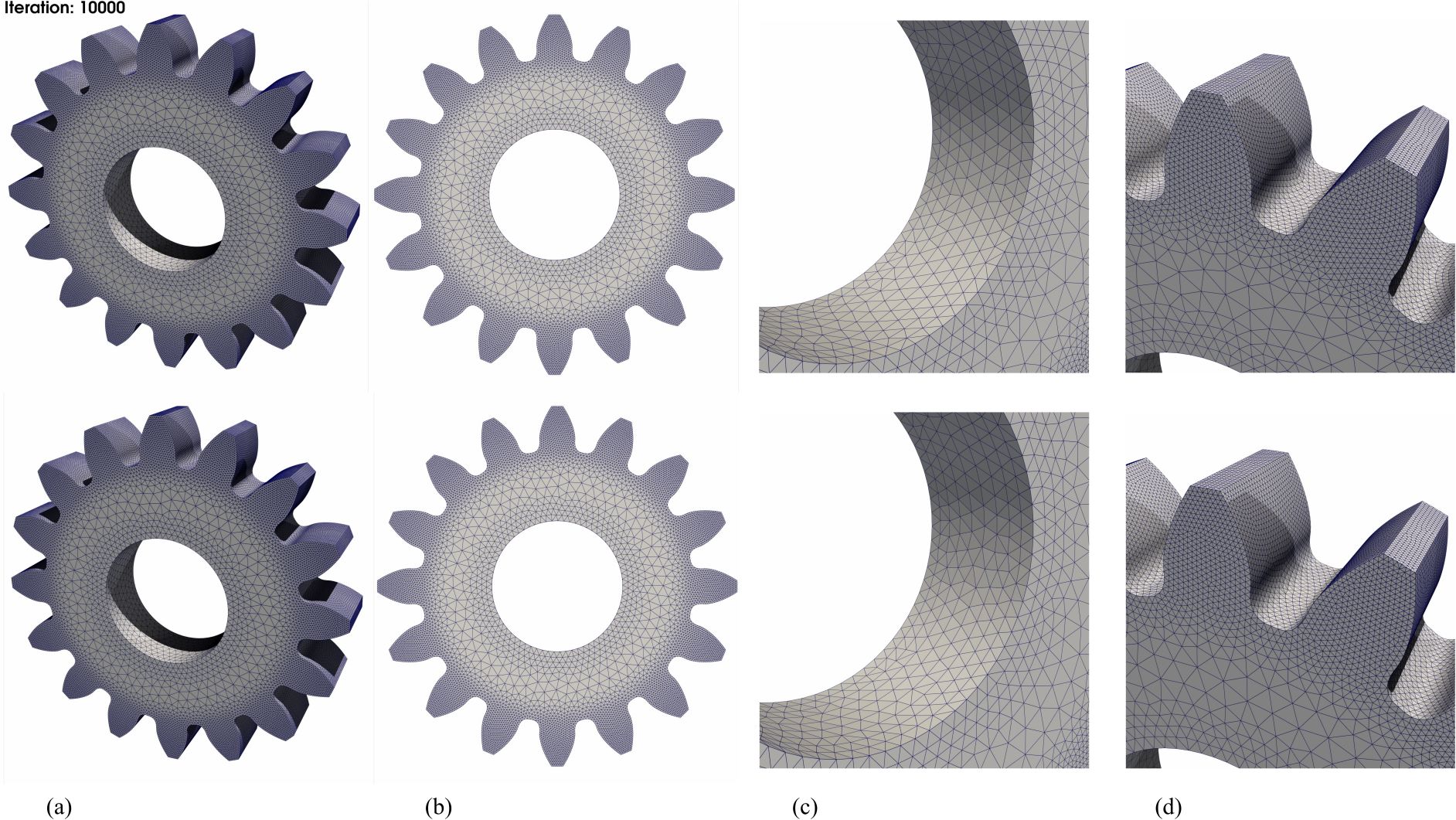}
		  \caption{Gear01: Delaunay triangulation of the final mesh. (a)(b) The full-sized views. (c)(d) The zoomed-in views at different camera positions. Upper row: results calculated by the algorithm developed by Fu et al. \cite{FU2019396}. Bottom row: results calculated by the proposed algorithm.}
		\label{fig:gear_mesh}
		\end{figure}

		The initial condition of ``Gear01" is plotted in Fig. \ref{fig:gear_stats} (a). From the time histories of $\theta_{<30}$, $G_{avg}$, and $G_{min}$ (see Fig. \ref{fig:gear_stats} (b-d)), both methods achieves an equilibrium state. Similar mesh qualities are observed for both methods and 78\% reduction in wall-clock time is accomplished by the proposed method (see table \ref{Tab:validation_Gear01}). From the particles distributions at different instants (Fig. \ref{fig:gear_particle}) and the final triangulated mesh (Fig. \ref{fig:gear_mesh}), the ``\textit{improved}" method can maintain a stable simulation even when the accumulation of momentum is allowed and the particles of all features are evolved together. The geometry is fully recovered.

		\begin{table}[h]
		\centering
		\caption{Mesh quality of the Gear01 case}
		\scriptsize
		\label{Tab:validation_Gear01}
		\newcommand{\tabincell}[2]{\begin{tabular}{@{}#1@{}}#2\end{tabular}}
		\begin{tabular}{>{\centering\arraybackslash}m{1.2cm}
		                >{\centering\arraybackslash}m{0.6cm}
		                >{\centering\arraybackslash}m{0.6cm}
		                >{\centering\arraybackslash}m{1cm}
		                >{\centering\arraybackslash}m{1cm}
		                >{\centering\arraybackslash}m{1cm}
		                >{\centering\arraybackslash}m{1cm}
		                >{\centering\arraybackslash}m{1cm}
		                >{\centering\arraybackslash}m{1cm}
		                >{\centering\arraybackslash}m{1cm}}
		\hline
		 & $G_{avg}$ & $G_{min}$ & $\theta_{max}$ & $\theta_{min}$ & $\theta_{min}^{\#}$ & $\theta_{<30}$ & $N_{tri}$ & $runtime$ & $N_{iter}$\\ \hline
		 \textit{Fu et al.} & 0.94 &  0.49 &  116.68 & 31.29 & 55.17 & 0 & 85,292 & 473.77 & 7,900 \\
		 \textit{Improved}  & 0.94 &  0.53 &  112.26 & 33.59 & 55.06 & 0 & 85,286 & 104.17 & 1,600  \\
		\hline
		\end{tabular}
		\end{table}

\begin{figure}[H]
		  \centering
		    \includegraphics[width=0.8\textwidth]{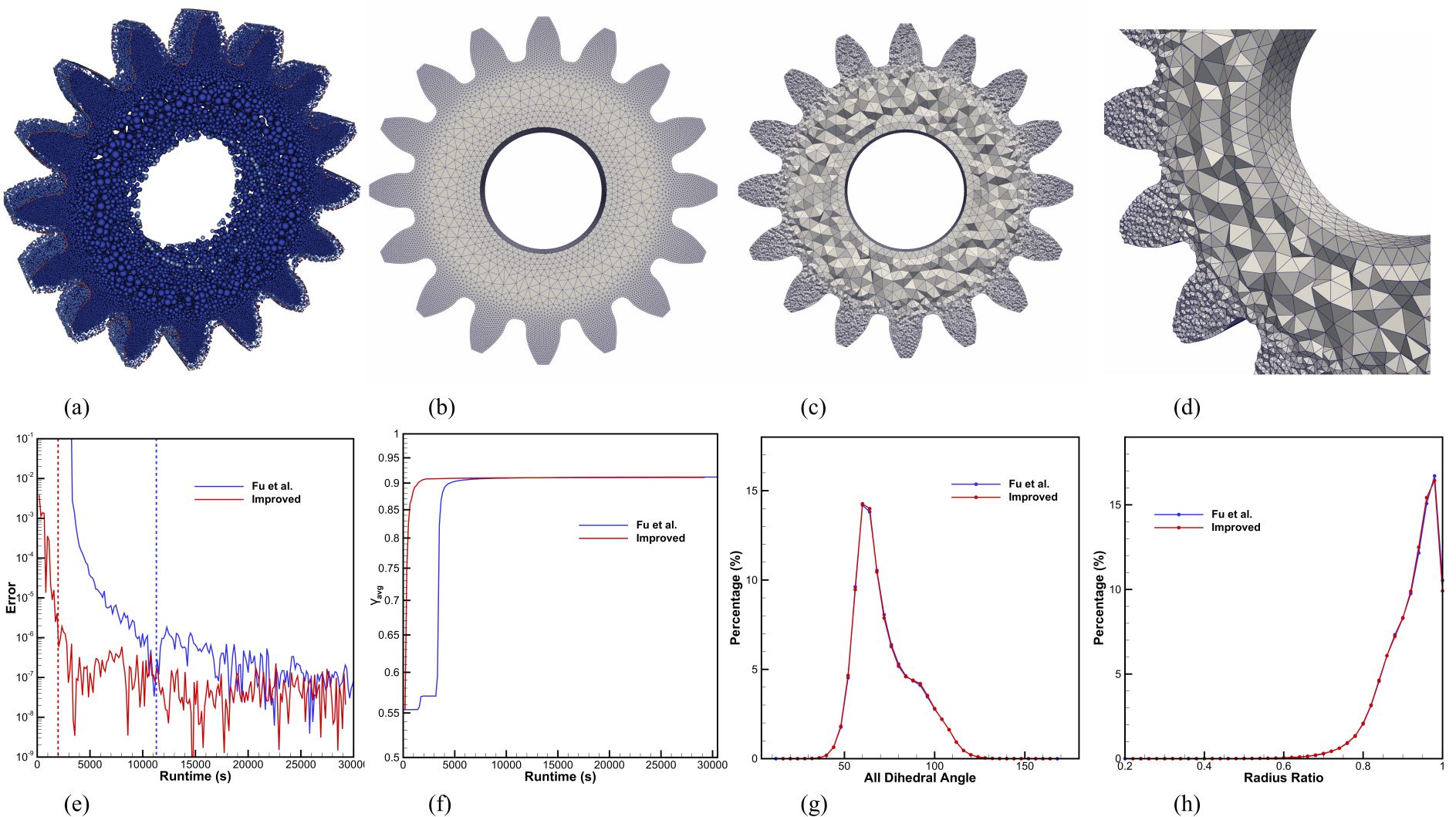}
		  \caption{Gear02: (a) Initial particle distribution. (b) Tetrahedralization of the final mesh with full sized view. (c) Tetrahedralization of the final mesh with clipped view. (d) Zoomed-in clipped view of the final mesh. The convergence histories of (e) $\overline{E}_{sys}$ and (f) $\gamma_{avg}$. Histogram of (g) the dihedral angle distribution and (h) the radius ratio distribution.}
		\label{fig:gear02_mesh}
		\end{figure}

		The initial condition of ``Gear02" is plotted in Fig. \ref{fig:gear02_mesh} (a). From the results illustrated in Fig. \ref{fig:gear02_mesh} (b-h), similar conclusions can be made with the previous case.

		\begin{table}[h]
		\centering
		\caption{Mesh quality of the Gear02 case}
		\scriptsize
		\label{Tab:validation_Gear02}
		\newcommand{\tabincell}[2]{\begin{tabular}{@{}#1@{}}#2\end{tabular}}
		\begin{tabular}{>{\centering\arraybackslash}m{1.2cm}
		                >{\centering\arraybackslash}m{1.4cm}
		                >{\centering\arraybackslash}m{1.4cm}
		                >{\centering\arraybackslash}m{0.5cm}
		                >{\centering\arraybackslash}m{0.5cm}
		                >{\centering\arraybackslash}m{0.5cm}
		                >{\centering\arraybackslash}m{0.5cm}
		                >{\centering\arraybackslash}m{0.8cm}
		                >{\centering\arraybackslash}m{0.8cm}
		                >{\centering\arraybackslash}m{0.8cm}
		                >{\centering\arraybackslash}m{0.8cm}}
		\hline
		 & $\theta_{min}/\theta_{max}$ & $\gamma_{min}/\gamma_{avg}$ & $\theta_{min}^{\#}$ & $\theta_{<10}$ & $\theta_{<20}$ & $\theta_{<30}$ & $\theta_{<40}$ & $N_{tet}$ & $runtime$ & $N_{iter}$\\ \hline
		 \textit{Fu et al.} & 14.24/159.16 &  0.23/0.91 & 55.86 & 0 & 6 & 293 & 10,534 & 884,095 & 9,664.05 & 12,200 \\
		 \textit{Improved}  & 12.54/160.93 &  0.22/0.91 & 55.74 & 0 & 13 & 349 & 11,170 & 884,118 & 2,187.22 & 2,600 \\
		\hline
		\end{tabular}
		\end{table}
\section{Conclusions}
\label{S:conclusion}

	In this paper, a feature-aware SPH formulation is proposed to achieve concurrent and automated isotropic unstructured mesh generation. Various numerical experiments are investigated considering both triangular surface-mesh and tetrahedral volumetric-mesh generation of geometries consisting various sharp features (creases, sharp edges and singularity points). The robustness of the particle relaxation procedure is improved and a remarkable speedup for obtaining an optimized mesh vertex distribution is accomplished. The main contribution of the paper can be summarized as:

	\begin{description}
		\item[1] Consistent cell-based feature definition system and target-information formulations are developed from previous work in \cite{FU2019396}\cite{ji2019consistent}. The new feature definition allows for more accurate characterization of complex geometries and particle systems;
		\item[2] A feature-aware correction term is introduced to the governing equations to resolve the issue of incomplete kernel support for particles close to the boundary vicinity. Intensive numerical validations demonstrate that the robustness of the simulation is improved. The constraint of nullifying particle momentum every timestep can be relaxed. Significantly improved convergence benefiting from the correction term is observed;
		\item[3] A new measurement of the convergence error, i.e. $\overline{E}_{sys}$, is proposed based on the concept of particle specific volume. $\overline{E}_{sys}$ is evaluated based on particle summation and no explicit geometric operations are needed. Numerical results show that the proposed criterion captures the convergence of the system successfully;
		\item[4] A two-phase mesh generation model is proposed to merge the initial mesh generation step and mesh-quality optimization step into the same set of governing equations characterized by different simulation parameters. Numerical experiments and results comparison reveal that the convergence of feature-size adaptation target is significantly improved and the mesh-quality-optimization step achieves the same performance as the original algorithm;
		\item[5] The proposed algorithm is compared with other state-of-the-art variation-based mesh generators. Comparable mesh-quality is obtained by the proposed method with significantly smaller computational costs.
	\end{description}

	In terms of future development, we will further improve the performance of the proposed method by harnessing the computational power of GPU-based architectures. Moreover, extending the proposed algorithm to anisotropic unstructured-mesh generation is of high interest.


\label{}



\section*{Acknowledgments}

Zhe Ji is partially supported by China Scholarship Council (No. 201506290038). Xiangyu Hu acknowledges funding Deutsche Forschungsgemeinschaft (HU1527/10-1 and HU1527/12-1).



  \bibliographystyle{elsarticle-num}
  \scriptsize
  \setlength{\bibsep}{0.5ex}

\bibliography{bib}

\end{document}